\documentclass[conference]{IEEEtran}
\IEEEoverridecommandlockouts
\usepackage{url} 
\usepackage{framed}
\usepackage{mdframed}
\usepackage{listings}
\usepackage{multicol}
\usepackage{amssymb}
\usepackage{lipsum}
\usepackage{adjustbox}
\usepackage[font=bf,skip=\baselineskip]{caption}
\usepackage{tabularx}
\usepackage{seqsplit}
\usepackage{multirow}
\usepackage{booktabs}
\usepackage{graphicx}
\usepackage{hyperref}
\usepackage{subcaption}
\usepackage{makecell}
\usepackage{xcolor}
\usepackage[ruled,vlined,linesnumbered] {algorithm2e}
\usepackage{amsmath}
\usepackage{tikz}
\usepackage{tcolorbox}
\usepackage{threeparttable}
\usepackage{tablefootnote}
\usepackage{csquotes}
\usepackage{colortbl}
\usepackage{algorithm2e}
\usepackage{adjustbox}
\usepackage{balance}
\usepackage{dblfloatfix}
\usepackage{placeins}
\usepackage{afterpage}

\usepackage{breakurl} 
\usepackage{boldline}
\usepackage[ruled,vlined]{algorithm2e}
 
\usepackage{pifont}
\usepackage{balance}
\let\labelindent\relax
\usepackage{enumitem}
\usepackage[htt]{hyphenat}

\definecolor{customblue}{HTML}{006ca6}
\definecolor{customgreen}{HTML}{009264}
\definecolor{custombrown}{HTML}{ff3d00}
\AtEndPreamble{
 \usepackage{hyperref}
 \hypersetup{
  colorlinks = true,
  linkcolor = customgreen,
  anchorcolor = purple,
  citecolor = customgreen,
  filecolor = purple,
  urlcolor = customblue
 }
}
\definecolor{customcell}{HTML}{ffebb7}

\newcommand{\yes}{\tikz\draw[fill=black] (0,0) circle (.4em);}
\newcommand{\no}{\tikz\draw[fill=white] (0,0) circle (.4em);}
\newcommand{\partialcell}{\begin{tikzpicture}
    \filldraw[fill=white] (0,0) circle (.4em);
    \filldraw[fill=black] (0,.4em) arc (90:270:.4em);
\end{tikzpicture}}

\newcommand{\find}[1]{
\begin{tcolorbox}[leftrule=0.5mm,toprule=0mm,bottomrule=0mm,left=0.7pt,right=0.7pt,top=0.2pt,bottom=0.2pt]
\em #1
\end{tcolorbox}
}

\newcommand{\ea}{\textit{et~al.}}
\newcommand{\toxicdict}{\textit{ToxicDict}}

\begin{document}

\title{On the (In)Security of LLM App Stores}

\author{
\IEEEauthorblockN{Xinyi Hou\IEEEauthorrefmark{1}, Yanjie Zhao\IEEEauthorrefmark{1}, and Haoyu Wang\IEEEauthorrefmark{2}}
\IEEEauthorblockA{
Huazhong University of Science and Technology, Wuhan, China\\
xinyihou@hust.edu.cn, yanjie\_zhao@hust.edu.cn, haoyuwang@hust.edu.cn}
\thanks{\IEEEauthorrefmark{1}Xinyi Hou and Yanjie Zhao contributed equally to this work.}
\thanks{\IEEEauthorrefmark{2}Haoyu Wang is the corresponding author (haoyuwang@hust.edu.cn).}
}

\maketitle

\begin{abstract}
LLM app stores have seen rapid growth, leading to the proliferation of numerous custom LLM apps. However, this expansion raises security concerns. In this study, we propose a three-layer concern framework to identify the potential security risks of LLM apps, i.e., LLM apps with abusive potential, LLM apps with malicious intent, and LLM apps with exploitable vulnerabilities. Over five months, we collected 786,036 LLM apps from six major app stores: GPT Store, FlowGPT, Poe, Coze, Cici, and Character.AI. Our research integrates static and dynamic analysis, the development of a large-scale toxic word dictionary (i.e., \toxicdict{}) comprising over 31,783 entries, and automated monitoring tools to identify and mitigate threats. We uncovered that 15,146 apps had misleading descriptions, 1,366 collected sensitive personal information against their privacy policies, and 15,996 generated harmful content such as hate speech, self-harm, extremism, etc. Additionally, we evaluated the potential for LLM apps to facilitate malicious activities, finding that 616 apps could be used for malware generation, phishing, etc. Our findings highlight the urgent need for robust regulatory frameworks and enhanced enforcement mechanisms.

\end{abstract}

\section{Introduction}
\label{sec:introduction}

Large Language Models (LLMs) such as ChatGPT~\cite{chatgpt}, Gemini~\cite{google2023gemini}, and Copilot~\cite{microsoft2023copilot} are at the forefront of the rapidly evolving LLM app store ecosystem. These platforms host a myriad of \textbf{custom LLM apps} that significantly enhance their functionality. Custom LLM apps are specialized apps built on top of general-purpose LLMs, designed to perform specific tasks or cater to particular domains by utilizing custom instructions, knowledge bases, and integrations with external services. These apps are hosted on \textbf{LLM app stores}~\cite{zhao2024llm}.
LLM app stores are experiencing a surge in popularity, as evidenced by platforms like FlowGPT~\cite{flowgpt} with its 4 million monthly active users and recent \$10 million funding~\cite{pr2024flowgpt}.

Unfortunately, the nascent stage of this development carries security concerns. For example, \textit{\textbf{instructions}} serve as the ``source code'' for LLM apps, allowing developers to dictate the behavior of these apps. If these instructions contain inappropriate content, such as jailbreaking prompts~\cite{chao2024jailbreakbench}, they can lead to malicious behavior by the LLM apps, adversely affecting users. In addition, malicious developers might intentionally upload harmful \textit{\textbf{knowledge files}} or integrate malicious \textit{\textbf{third-party services}} to exploit the powerful capabilities of LLM apps for nefarious activities such as generating malware code or crafting phishing emails.

Recent OpenAI threat reports~\cite{openai2024disrupting} have highlighted several instances of LLM misuse over the past three months, underscoring the significant threat that exists within LLM app ecosystems.
Despite the implementation of various policies~\cite{bytedance2024terms,ciciterms,openai2024usage,shazeer2024terms} aimed at regulating LLM app behavior, these policies are often vague and not rigorously enforced. Prominent platforms like OpenAI~\cite{openai2024usage} and Coze~\cite{bytedance2024terms} claim to conduct regular reviews of LLM apps in their app stores and promptly remove those that violate their policies. These review mechanisms include OpenAI's \texttt{Moderations}~\cite{moderations} endpoint, red teaming~\cite{red_teaming} methods, etc. During our five-month crawl of LLM apps, we observed that \textbf{5,462 apps were removed after a certain period, 132 of these removals were likely due to policy violations}. 
Consider the app with ID ``g-1vQR4hP8T'' from OpenAI's GPT Store~\cite{GPTStore} as an illustrative example. This app was removed for dispensing medical advice, an action that contravenes OpenAI's usage policies.

Despite these measures,  the overwhelming number of LLM apps in popular stores poses a substantial challenge for platform administrators. For example, with GPT Store hosting over three million LLM apps~\cite{openai2024gptstore} and FlowGPT housing hundreds of thousands~\cite{flowgpt10m}, the scale severely strains review processes.
In this paper, we examine six prominent LLM app stores, uncovering significant discrepancies in regulatory enforcement across platforms and highlighting critical security concerns within the LLM app ecosystem.
To the best of our knowledge, this is the first comprehensive and in-depth study examining the current state of LLM app store security. 
Previous research, notably Lin et al.'s~\cite{lin2024malla} empirical study on \textit{LLM-integrated malicious services}, has primarily focused on explicitly malicious paid LLM services, which are costly and limited in number. 
In contrast, we investigate \textit{LLM app stores}, where the development and usage costs of LLM apps are minimal, and the potential for widespread impact due to security vulnerabilities is substantial. Our objective is to shed light on the overlooked aspects of LLM app stores and conduct a thorough examination of their security landscape.

\begin{figure*}
    \centering
    \includegraphics[width=1\linewidth]{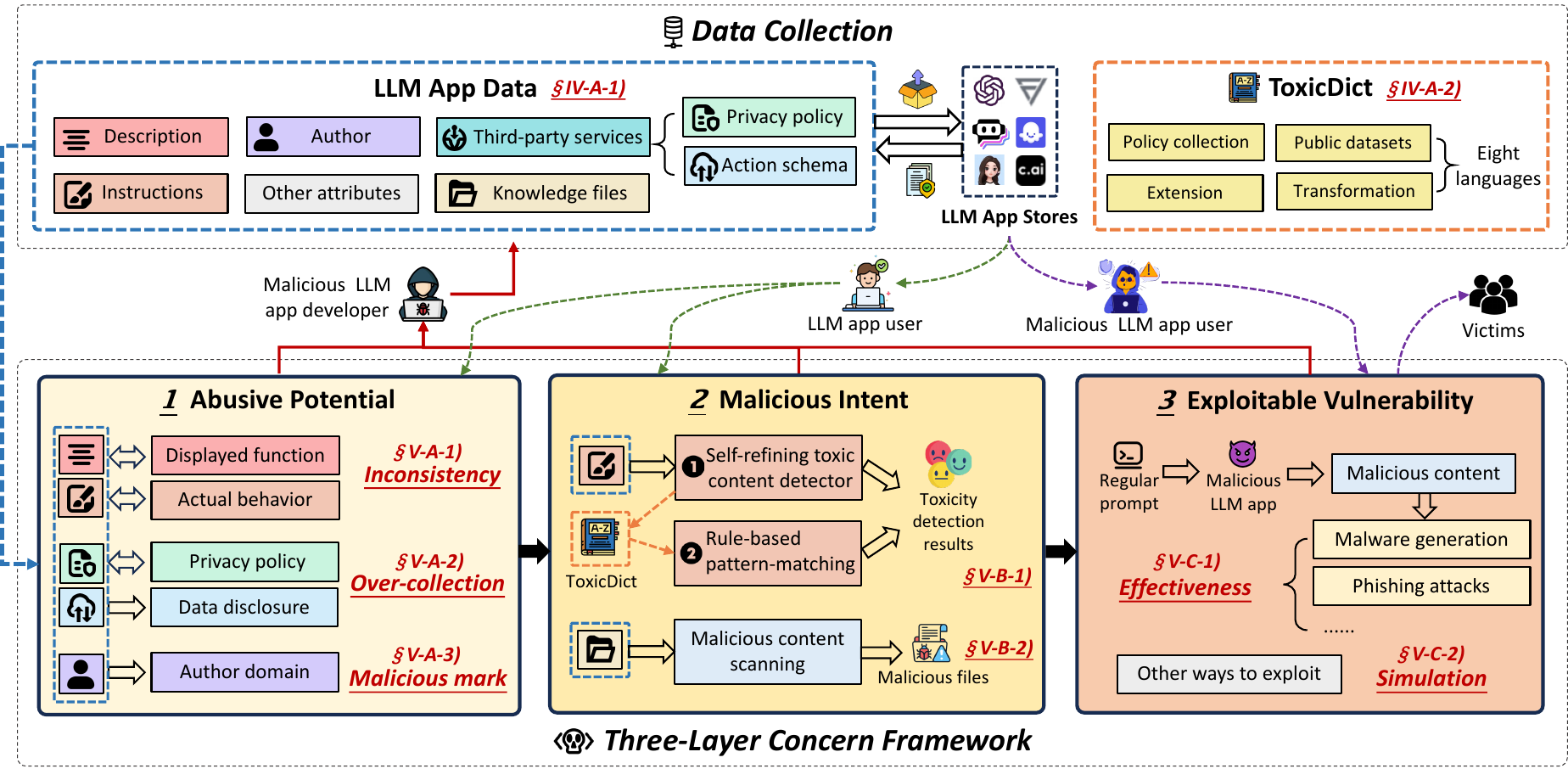}
    \caption{Overview of the three-layer security concern framework.}
    \label{fig:overview}
\end{figure*}

We propose a comprehensive three-layer concern framework, illustrated in~\autoref{fig:overview}, for the systematic analysis of LLM app security concerns.
The first layer, \textbf{LLM apps with abusive potential}, examines inconsistencies and potential misuse without definitive evidence of malicious content. This includes mismatched descriptions and instructions, improper data collection, suspicious author domains, etc. These issues primarily affect individual users who may be misled or have their data mishandled.
The second layer, \textbf{LLM apps with malicious intent}, focuses on apps specifically designed to harm users by directly embedding harmful functionalities. These apps pose an immediate threat to their users. The third layer, \textbf{LLM apps with exploitable vulnerabilities}, addresses apps containing malicious knowledge or flaws that can be exploited by attackers. These vulnerabilities have the potential to cause widespread security issues, impacting a broader range of victims beyond the app's immediate users.

To investigate and analyze these concerns, we developed an automated framework capable of detecting security risks from these three perspectives. 
Over five months, we crawled a total of 786,036 LLM apps from six LLM app stores: GPT Store~\cite{GPTStore}, FlowGPT~\cite{flowgpt}, Poe~\cite{poe}, Coze~\cite{coze}, Cici~\cite{cici}, and Character.AI~\cite{characterai}. Our study integrates both static and dynamic analysis, leveraging a large-scale toxic word dictionary (i.e.,\toxicdict{}) and automated tools for continuous monitoring. We discovered numerous instances where app descriptions did not match their instructions, potentially misleading users and hiding malicious intentions. We also identified apps that collected sensitive personal information in ways that contradicted their privacy policies. Furthermore, we categorized and detected harmful content such as hate speech, self-harm, and extremism, and evaluated the potential for LLM apps to execute malicious actions like malware generation and phishing. This approach provides real-time insights into emerging threats, enabling timely interventions to safeguard users.

\noindent\textbf{Contributions.} Our primary contributions\footnote{We will make our data and tools publicly available upon acceptance.} are as follows:
\begin{enumerate}

    \item Our research presents the first comprehensive empirical study of security concerns in LLM app stores. We propose a novel three-layer concern framework for LLM app security analysis, encompassing LLM apps with abusive potential, LLM apps with malicious intent, and LLM apps with exploitable vulnerabilities.

    \item To facilitate our analysis, we developed an automated framework that combines static and dynamic approaches. The static analysis utilizes \toxicdict{}, our custom-built dictionary containing 31,783 toxic words across 14 categories in eight languages, enabling effective preliminary detection of toxic content. Our framework also incorporates dynamic interaction with LLM apps to identify their actual behavior, complemented by automated tools that provide continuous monitoring of app stores.

    \item We collected 786,036 LLM apps from six different stores. Our investigation of these apps revealed widespread security issues, including 16,376 apps with abusive potential, 15,996 apps with malicious intent, and 616 apps with exploitable vulnerabilities.

\end{enumerate}

\section{Background}
\label{sec:background}
\subsection{LLM App Store}

The rapid development of LLMs has propelled the growth of a series of downstream applications, such as LLM app stores, on-device LLMs, and expert domain-specific LLMs~\cite{wang2024large}. Among these, LLM app stores have emerged as prominent centralized platforms for hosting and distributing custom LLM-powered applications. These stores offer a diverse array of intelligent services tailored to various purposes, tasks, and scenarios, allowing users to easily discover and access LLM apps~\cite{zhao2024llm}. While the LLM app ecosystem has unlocked tremendous potential for innovation and efficiency, it also presents opportunities for malicious actors to exploit LLM capabilities for harmful purposes.

Several factors contribute to the security challenges of LLM app stores. \textbf{The low barrier to entry for creating LLM apps} enables individuals with minimal technical expertise to develop and deploy potentially malicious apps, a problem exacerbated by inadequate vetting processes in some stores. Additionally, \textbf{the ability to integrate external knowledge sources and third-party services} opens avenues for exploitation by malicious actors who can spread disinformation, propagate scams, or compromise user privacy.
The security risks are further amplified by \textbf{the ability of LLMs to generate highly convincing content}. This capability allows for the creation of apps that produce fake news, impersonate legitimate entities, or manipulate public opinion with alarming effectiveness. Moreover, \textbf{the lack of comprehensive monitoring and enforcement mechanisms} in LLM app stores, combined with the high volume and rapid pace of app development, makes it challenging to promptly identify and remove malicious apps.

\subsection{Policy Regulations}

To address the challenge of ensuring compliance amidst the rapid growth of LLM apps, each LLM app store has established clear policies to regulate the development process. These policies outline the guidelines and restrictions developers must follow when creating and publishing their apps on their respective platforms. As shown in~\autoref{tab:policy}, the policies typically cover three main aspects:

\begin{itemize}
    \item \textbf{Privacy policy} informs users about the data collection and usage practices of the app. While most LLM app stores have detailed privacy policies~\cite{bytedance2024privacy,ciciprivacy,openai2024privacy,quora2024privacy,shazeer2024privacy}, some like FlowGPT~\cite{flowgptprivacy} have incomplete policies that require further refinement.
    \item \textbf{Usage guidelines} help developers create and maintain apps~\cite{openai2024usage,quora2024usage}. Although FlowGPT~\cite{flowgptcontent} and Character.AI~\cite{shazeer2024guidelines} have guidelines, their content is simplistic. Some LLM app stores, like Coze and Cici, lack guidelines entirely, highlighting the need for comprehensive policies.
    \item \textbf{Terms of service} outlines the legal agreements between the app store and users. Notably, all the LLM app stores examined have terms of service in place~\cite{bytedance2024terms,ciciterms,openai2024terms,quora2024terms,shazeer2024terms,flowgptterms}.
\end{itemize}

\begin{table}[h!]
\centering
\caption{LLM app stores and their policy regulations.}
\resizebox{1\linewidth}{!}{
\begin{threeparttable}{
\begin{tabular}{cccc}
\toprule[1.2pt]
\textbf{Store name} & \textbf{Privacy policy} & \textbf{Usage guideline} & \textbf{Terms of service} \\ \midrule
GPT Store    & \yes & \yes & \yes \\ 
FlowGPT      & \partialcell & \partialcell & \yes \\ 
Poe          & \yes & \yes & \yes \\ 
Coze         & \yes & \no & \yes \\ 
Cici         & \yes & \no & \yes \\ 
Character.AI & \yes & \partialcell & \yes \\ 
\bottomrule[1.2pt]
\end{tabular}}
\begin{tablenotes}
    \footnotesize
    \item \yes\ indicates detailed policy, \partialcell\ indicates incomplete policy, \no\ indicates the absence of policy.
\end{tablenotes}
\end{threeparttable}}
\label{tab:policy}
\end{table}

LLM app stores employ both automated and manual review processes to enforce policies, using techniques like machine learning-based moderation~\cite{moderations} and red teaming~\cite{red_teaming}. However, they still face challenges in comprehensively identifying and mitigating malicious apps due to rapid development and content complexity. Malicious developers often exploit these challenges to circumvent moderation mechanisms.
Additionally, unlike conventional apps that typically provide their own privacy policies detailing permissions, data collection, and usage~\cite{slavin2016toward,wang2018guileak}, LLM app developers often only provide privacy policies of third-party platforms when used. This leaves users uncertain about how their data is being handled within the LLM app itself, highlighting a gap in transparency and user protection in the LLM app ecosystem.

\subsection{Threat Model}

\noindent\textbf{Assumptions and threat scenarios.}
As shown in~\autoref{fig:overview}, our three-layer concern framework encompasses various LLM app threat scenarios. We assume that these scenarios exist in LLM app stores. First, for LLM apps with abusive potential, we posit that some developers create apps with inconsistent descriptions or improper data practices, exploiting inadequate app store oversight. These primarily affect individual users through privacy violations and misunderstandings. Second, regarding LLM apps with malicious intent, we assume developers may intentionally design apps to generate harmful content or enable illegal activities, posing direct threats to users and potential broader societal harm. Finally, for LLM apps with exploitable vulnerabilities, we assume that LLM apps may contain vulnerabilities that malicious actors can leverage for various attacks, including malware generation, phishing, data theft, service disruption, and disinformation propagation. We further assume that these vulnerabilities can have far-reaching consequences beyond immediate users, potentially resulting in significant financial, reputational, and societal damage.

\noindent\textbf{Our goal.}
The primary goal of this study is to illuminate the security concerns prevalent in LLM app stores. Through an in-depth analysis of popular stores and their hosted apps, we aim to uncover hidden risks in this growing ecosystem. Our objectives include identifying and categorizing LLM app security issues, evaluating current regulatory measures, and proposing risk mitigation strategies for insecure LLM apps.

\section{Deﬁnitions}
\label{sec:definition}

\begin{table*}[h!]
\centering
\caption{Composition of data collected from LLM app stores.}
\resizebox{\linewidth}{!}{
\begin{threeparttable}
\begin{tabular}{cccccccccccccccc}
\toprule[1.2pt]
\textbf{Store name} & \textbf{LLM app ($A$)} & \multicolumn{2}{c}{\textbf{Description}} & \multicolumn{2}{c}{\textbf{Author}} & \multicolumn{2}{c}{\textbf{Instructions}} & \multicolumn{2}{c}{\textbf{Knowledge files}} & \multicolumn{3}{c}{\textbf{Third-party services}} & \textbf{Visibility}\tnote{1} \\
\cmidrule(lr){3-4} \cmidrule(lr){5-6} \cmidrule(lr){7-8} \cmidrule(lr){9-10} \cmidrule(lr){11-13}
 & \textbf{\# $A$}  & \textbf{\# $A$} & \textbf{\% $A$} & \textbf{\# $A$} & \textbf{\% $A$} & \textbf{\# $A$} & \textbf{\% $A$} & \textbf{\# $A$} & \textbf{\# files} & \textbf{\# $A$} & \textbf{\# Policy} & \textbf{\# Schema} &  \\
\midrule
GPT Store & 663,119 & 630,420 & 95.07\% & 241,621 & 36.44\% & 22,961 & 3.46\% & 45,690 & 192,714 & 5,498 & 6,547 & 5,767 & \yes~\partialcell~\no \\
FlowGPT & 34,345 & 34,339 & 99.98\% & 9,374 & 27.29\% & 24,983 & 72.74\% & 0 & 0 & / & / & / & \yes~\no \\
Poe & 16,544 & 16,050 & 97.01\% & 8,728 & 52.76\% & 6,063 & 36.65\% & 0 & 0 & / & / & / & \yes~\no \\
Coze & 51,918 & 19,666 & 37.88\% & 33,606 & 64.73\% & 1,491 & 2.87\% & 0 & 0 & / & / & / & \yes \\
Cici & 13,060 & 13,060 & 100.00\% & 9,468 & 72.50\% & 0 & 0.00\% & / & / & / & / & / & \yes~\partialcell~\no \\
Charcter.AI & 7,050 & 7,050 & 100.00\% & 6,252 & 88.68\% & 1,819 & 25.80\% & / & / & / & / & / & \yes~\partialcell~\no \\
\midrule
\textbf{Total} & 786,036 & 720,585 & 91.67\% & 309,049 & 39.32\% & 57,317 & 7.29\% & 45,690 & 192,714 & 5,498 & 6,547 & 5,767 & / \\
\bottomrule[1.2pt]
\end{tabular}
\begin{tablenotes}
    \footnotesize
    \item[1] \yes\ indicates public, \partialcell\ indicates workspace-specific~\cite{openai2024gptstore} (only visible to specific users), \no\ indicates private. 
    \item[2] ``/'' indicates the platform does not support this functionality.
\end{tablenotes}
\end{threeparttable}}
\label{tab:dataset}
\end{table*}

An \emph{LLM app} $A$ is defined as a tuple:
\begin{equation*}
A = (M, K, [S])
\end{equation*}

where:

\begin{itemize}
\item $M$ is the \emph{metadata} of the app, which includes elements such as the app's name, author, id, description, instructions, and other metadata:
\begin{equation*}
M = {m_1, m_2, \ldots, m_n} 
\end{equation*}
\begin{equation*}
= {\mathit{name}, \mathit{author}, \mathit{ID}, \mathit{description}, \mathit{instructions}, \ldots}
\end{equation*}
\item $K$ is the set of \emph{knowledge files} associated with the app:
\begin{equation*}
K = \{k_1, k_2, \ldots, k_m\}
\end{equation*}
\item $[S]$ is an optional \emph{third-party service integration}. If the app uses a third-party service, it must provide a schema $sc$ describing the data collected and the detailed usage of the service, as well as a privacy policy $pp$:
\begin{equation*}
S = (sc, pp)
\end{equation*}
\end{itemize}
The \emph{visibility scope} $V$ of the app, which can be public, workspace-specific (visible to team members or users with the link) or private, is defined as:
\begin{equation*}
V \in {\mathit{public}, \mathit{workspace}, \mathit{private}}
\end{equation*}
The \emph{LLM app store} has a set of policies $\Pi$ that govern the development and regulation of LLM apps:
\begin{equation*}
\Pi \owns \mathcal{A}
\end{equation*}
where $\mathcal{A}$ represents the set of all LLM apps in the LLM app store, and $\owns$ denotes that the policies $\Pi$ encompass and regulate the LLM apps in $\mathcal{A}$. The $\Pi$ consist of various components, including the \emph{usage guidelines} $U$, \emph{privacy policy} $P$, \emph{terms of service} $T$, and potentially other policies:
\begin{align*}
\Pi &\supset U, P, T, \ldots
\end{align*}
and all LLM apps must adhere to the $U$, $P$, $T$, and possibly other unnamed policies.

\section{Methodology}
\label{sec:methodology}

The methodology is structured into several key components. \textit{\textbf{A. Data Collection}} involves gathering data from LLM app stores and constructing our \toxicdict{}. \textit{\textbf{B. Detection of LLM Apps with Abusive Potential}} includes inconsistency analysis and malicious domain detection to identify potential abuse. \textit{\textbf{C. Detection of LLM Apps with Malicious Intent}} employs a self-refining toxic content detector and rule-based pattern matching to identify harmful intentions. Finally, \textit{\textbf{D. Verification of LLM Apps with Exploitable Vulnerability}} involves evaluating malicious behavior and hypothesizing malicious scenarios to verify vulnerabilities.

\subsection{Data Collection}
\label{sec:data_collection}
\noindent\textbf{\textit{1) LLM apps data collection}}

In the initial phase of our study, we systematically collected data from various LLM app stores known for hosting customized LLM apps. Our primary data sources included GPT Store~\cite{GPTStore}, FlowGPT~\cite{flowgpt}, Poe~\cite{poe}, Coze~\cite{coze}, Cici~\cite{cici}, and Character.AI~\cite{characterai}. To efficiently gather data from these sources, we developed an automated web scraping tool using Selenium~\cite{selenium}.~\autoref{tab:dataset} shows the composition of the data we collected from each LLM app store. Each platform's LLM app has a unique ID. Therefore, we use the ID as the identifier for LLM apps to count the number and serve as a reference.

\begin{itemize}
    \item \textbf{GPT Store}: We utilized a recently released dataset, GPTZoo~\cite{hou2024gptzoo}, which contains metadata for 730,420 LLM apps. Due to the lack of direct information on instructions, knowledge files, and third-party services in the OpenAI GPT Store, we had to employ reverse engineering to obtain data on instructions and knowledge files. To comply with OpenAI's usage policies, this reverse engineering process had to be conducted under a restriction on the number of interactions allowed, making it a highly time-consuming endeavor. To date, we have collected instructions for 22,961 LLM apps and found 45,690 apps including knowledge files. Additionally, using the Free GPTs Scraper~\cite{gptscraper} and the GPT Store's API endpoint, we have gathered information on third-party services usage for 182,697 LLM apps, successfully obtaining 5,767 action schemas for 5,498 of these LLM apps.

    \item \textbf{FlowGPT}: The homepage of FlowGPT displays detailed categories of LLM apps. By traversing all categories on the homepage using the FlowGPT API endpoint, we obtained specific information for 34,345 LLM apps. FlowGPT allows developers to choose whether to publicly share instructions with users, and we ultimately obtained instructions for 24,983 LLM apps.

    \item \textbf{Poe}: We used an automated tool to scrape the basic information of all categories of LLM apps from Poe, totaling 16,544 apps. We also checked each LLM app's page to see if instructions were publicly available, ultimately obtaining 6,063 sets of instructions.
    
    \item \textbf{Coze}: Coze offers two versions: one for mainland China and one for global use, with domains ending in \texttt{.cn} and \texttt{.com}, respectively. The LLM apps available on these two versions are not entirely the same. We scraped basic information for a total of 51,918 LLM apps from both versions of the store, but only 1,491 of these apps publicly provided instructions. Additionally, Coze supports the seamless integration of third-party plugins from its plugin store during the development of LLM apps, without the need to provide third-party privacy policies.
    
    \item \textbf{Cici}: Cici is a popular platform that primarily features virtual character LLM apps and supports switching between fifteen languages. However, the information available from Cici's LLM apps is quite limited, as creating an LLM app on Cici requires only a name and description. We collected metadata for a total of 13,060 LLM apps.
    
    \item \textbf{Character.AI}: Character.AI is also an LLM app store primarily featuring virtual character apps and supporting voice interactions. Similar to the GPT Store's display method, Character.AI does not fully showcase all categories of LLM apps. Therefore, we had to scrape LLM apps by searching with keywords and saving the search results. To focus our investigation on the security aspects of LLM apps in LLM app stores, we selected 232 keywords from our \toxicdict{} categorization to use as search terms. This approach allowed us to scrape a total of 7,050 LLM apps and 1,819 publicly available instructions.
\end{itemize}

To ensure the integrity and usability of our dataset, we undertook several preprocessing steps. Initially, we cleaned the data to remove incomplete, irrelevant, or duplicate entries. We then standardized the data formats across all platforms, ensuring consistency in metadata representation. This involved normalizing key attributes such as ID, description, author, instructions, knowledge files, and third-party service information. Additionally, we integrated third-party service data where applicable. Other attributes were retained as supplementary information for future experiments. Finally, we conducted thorough quality assurance checks to verify the accuracy and completeness of the processed data. 

\noindent\textbf{\textit{2) Construction of }\toxicdict{}}

Considering the limited scope of currently available public toxic word lists, we constructed a comprehensive dictionary, \toxicdict{}, which encompasses 31,783 toxic words across 14 categories. These categories include: 

\begin{displayquote}
\noindent\textit{Hate, Self-Harm, Sexual, Violence, Profanity, Extremism, Spam, Minors, Regulated, Personal Decisions, PII, Links, Gambling, and Political}. 
\end{displayquote}

\begin{figure}[h!]
    \centering
    \includegraphics[width=1\linewidth]{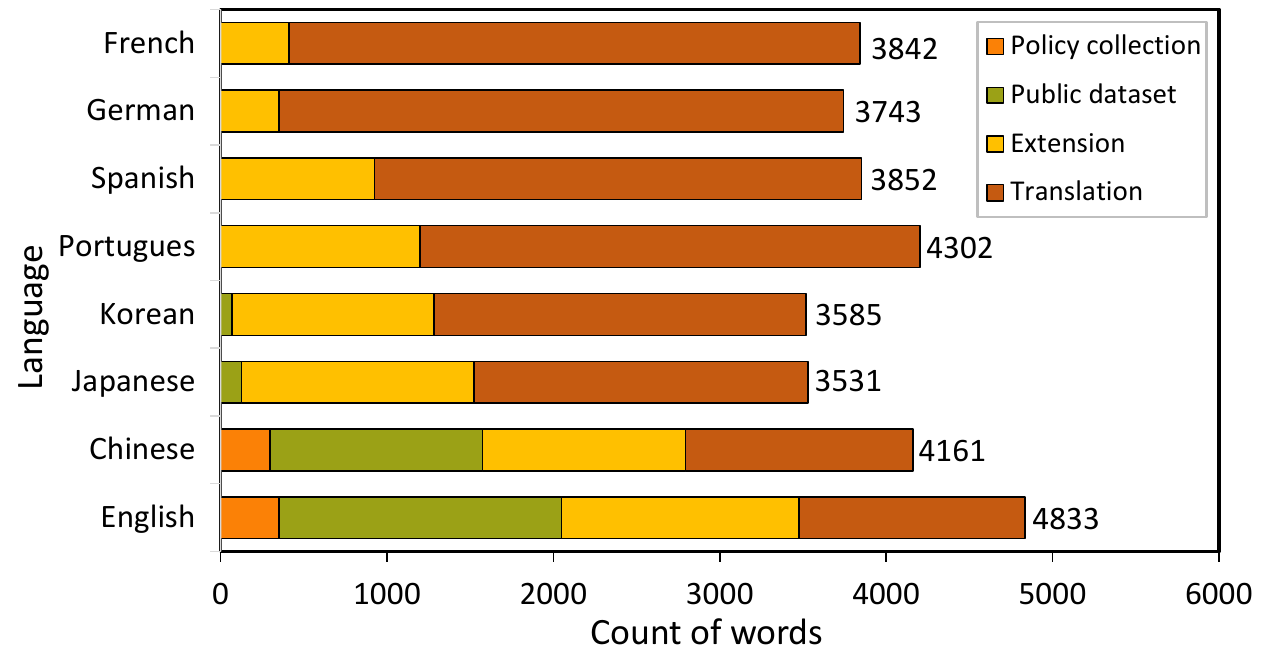}
    \caption{Language and source distribution of words in our \toxicdict{} dictionary.}
    \label{fig:toxicdict}
\end{figure}

\noindent The selection of these categories was informed by the policies of LLM app stores and the OpenAI \texttt{Moderation} endpoint~\cite{moderations}, ensuring comprehensive coverage of toxic content types, from hate speech and self-harm to privacy violations and spam. ~\autoref{fig:toxicdict} illustrates the distribution of languages and sources of the words in \toxicdict{}. The dictionary includes words from eight languages, selected based on their prevalence among LLM apps in the GPT store~\cite{GPTStore}. 
In detail, the sources of toxic words include:

\begin{itemize}
    \item \textbf{Policy collection}:
    We extracted toxic words from privacy policies, usage guidelines, and terms of service of LLM app stores. This ensures our \toxicdict{} reflects content explicitly prohibited by these platforms, aiding in identifying LLM app violations and potential misuse.

    \item \textbf{Public dataset}: We incorporated words from established public datasets
    on platforms like GitHub~\cite{chamoli2022text,grad2021orthrus,patch2020bad} and Hugging Face~\cite{lin2023toxicchat,toxicconversations}, providing a foundational set of known harmful or inappropriate terms.

    \item \textbf{Extension}: We utilized the powerful language capabilities of GPT-4o~\cite{gpt4o} to expand our existing word lists, identifying and generating additional toxic words that fit within our defined categories.
    
    \item \textbf{Translation}: To cover a broader range of languages, we translated toxic words from English and Chinese into other languages using GPT-4o. During the translation process, we instructed the GPT-4o to retain the linguistic characteristics and nuances of each target language as much as possible.

\end{itemize}  

\subsection{Detection of LLM Apps with Abusive Potential}

\noindent\textbf{\textit{1) Inconsistency analysis}}

\noindent\textbf{Content inconsistency.}
We developed a consistency analysis tool based on Llama3-8B~\cite{llama3}, as shown in~\autoref{alg:consistency_analysis_tool}, which takes the description and instructions of LLM apps as input.
The tool assesses consistency between description and instructions, considering relevance, detail alignment, and task coherence. It assigns a consistency score from 0 to 1, where 0 indicates completely different tasks and 1 signifies instructions that precisely extend the description. The tool also provides a rationale for the score to aid analysis.
The output is typically in JSON format, including fields like \textit{id}, \textit{consistency\_score}, and \textit{reason}. If the tool fails to produce a correct output, it attempts the check up to three times. Persistent errors are flagged for external review. After detection, reasons are categorized, and an analysis summary of inconsistencies is provided. This analysis is crucial for auditing potential misuse, as inconsistencies can mislead users and hide malicious intent.

\begin{algorithm}[h!]
\caption{Consistency Analysis Tool}
\label{alg:consistency_analysis_tool}
\footnotesize
\SetKwInput{KwInput}{Input}
\SetKwInput{KwOutput}{Output}
\DontPrintSemicolon

\KwInput{LLM app dataset $D$, Consistency analysis model $M$}
\KwOutput{Set of LLM apps with inconsistency $S$, Summary of inconsistency analysis}

$S \gets \emptyset$\;

\ForEach{LLM app $A \in D$}{
    Extract \textit{id}, \textit{description}, \textit{instructions} from $A$\;
    $P \gets \text{Construct\_Prompt}(\textit{id}, \textit{description}, \textit{instructions})$\;
    
    \For{$attempt \gets 1$ \KwTo $3$}{
        $O \gets M(P)$\;
        $(\textit{consistency\_score}, \textit{reason}) \gets \text{Extract\_Results}(O)$\;

        \If{$\textit{consistency\_score} \neq \text{None}$}{
            \textbf{break}\;
        }
    }

    \If{$\textit{consistency\_score} = 0$}{
        $\textit{consistency\_score} \gets \text{``Requires external feedback''}$\;
        $\textit{reason} \gets \text{``Manual review needed''}$\;
    }
    
    \If{$\textit{consistency\_score} < \text{threshold}$}{
        $S \gets S \cup \{(A, \textit{consistency\_score}, \textit{reason})\}$\;
    }
}

$\text{categories} \gets \text{Categorize\_Reasons}(S)$\;
$\text{summary} \gets \text{Generate\_Summary}(S, \text{categories})$\;

\Return $S$, $\text{summary}$\;
\end{algorithm}

\noindent\textbf{Data type inconsistency.} To analyze the data types collected by third-party services from the Action schema~\cite{gptaction}, we extracted relevant information using natural language processing (NLP) techniques. Our goal is to uncover potential LLM app abuse, particularly focusing on sensitive data types that could be misused for profiling users or targeted advertising. We parsed the Action schema JSON files to list the data types collected by third-party services. Using NLP, we normalized and categorized these data types, creating a comprehensive list. We then cross-referenced this list with 32 sensitive data types identified from LLM app store privacy policies. These sensitive data types include personal identifiers, location data, conversation history, etc.
To assess the consistency between the collected data and the declared data collection practices, we used \texttt{Polisis}~\cite{polisis} to analyze the privacy policies of LLM app stores. \texttt{Polisis} automatically detects and categorizes data practices, allowing us to compare the data types declared in the privacy policies with those actually collected, as stated in the Action schema.

\noindent\textbf{\textit{2) Malicious domain detection}}

Some LLM app developers publicly disclose their domain, referred to as the author domain. To ensure the safety and legitimacy of these domains, we utilize tools such as VirusTotal~\cite{virustotal} and Google Safe Browsing~\cite{safebrowsing} to scan these domains for any malicious activity. VirusTotal aggregates many antivirus products and online scan engines to check for viruses, worms, trojans, and other kinds of malicious content detected in the scanned domains. Google Safe Browsing provides lists of URLs for web resources that contain malware or phishing content, which is regularly updated and used to protect users from unsafe web content.
If an author domain is flagged as malicious by these tools, it implies that the developer associated with this domain may have malicious intent or has been compromised. This could potentially mean that the LLM app itself is being used to disseminate harmful content or engage in other abusive activities. Similarly, we can perform scans on action domains, which are the domains associated with third-party services used by the LLM app. By scanning these domains, we can detect the presence of potentially malicious third-party services integrated into the LLM app. Malicious domain detection helps uncover LLM apps with abusive potential by identifying domains that are linked to known malicious activities. 

\subsection{Detection of LLM Apps with Malicious Intent}
\label{sec:malicious_intent}
We use a complementary approach to achieve comprehensive and efficient detection of harmful content.
\textit{Self-refining LLM-based toxic content detector} considers context and cultural nuances, improving intent discernment and prediction accuracy.
It continuously learns from new instances and updates the \toxicdict{}, thereby enhancing the accuracy and coverage of \textit{rule-based pattern matching}.
The rule-based method provides immediate, targeted detection results, compensating for the precision limitations of LLM-based approaches.

\begin{algorithm}[h!]
\caption{Self-refining Toxic Content Detector}
\label{alg:self_refining_toxic_content_detector}
\footnotesize
\SetKwInput{KwInput}{Input}
\SetKwInput{KwOutput}{Output}
\DontPrintSemicolon

\KwInput{LLM app dataset $D$, LLM-based toxic content detector $M$}
\KwOutput{Set of LLM apps with toxic content $T$, Summary of toxic content analysis}

$H \gets \emptyset$ // $H$ is the set of challenging instances \;
$T \gets \emptyset$\;

\ForEach{LLM app $A \in D$}{
    Extract \textit{id}, \textit{instructions} from $A$\;
    $P \gets \text{Construct\_Prompt}(\textit{id}, \textit{instructions})$\;
    
    $O \gets M(P)$\;
    $(\textit{toxicity\_scores}, \textit{toxic\_words}) \gets \text{Extract\_Results}(O)$\;

    \If{$\textit{toxicity\_scores} = \text{None}$}{
        $H \gets H \cup \{A\}$\;
    } \Else {
        $T \gets T \cup \{(A, \textit{toxicity\_scores}, \textit{toxic\_words})\}$\;
    }
}

\If{$|H| > 0$}{
    $\text{sampled\_challenging\_instances} \gets \text{Random\_Sample}(H, 10)$\;
    \text{Manual\_Review}($\text{sampled\_challenging\_instances}$)\;
    \ForEach{$instance \in \text{sampled\_challenging\_instances}$}{
        $\text{Update\_Model}(M, instance)$\;
    }
}

$\text{summary} \gets \text{Generate\_Summary}(T)$\;

\Return $(T, \text{summary})$\;
\end{algorithm}

\noindent\textbf{\textit{1) Self-refining LLM-based toxic content detector}}

The self-refining LLM-based toxic content detector leverages the advanced capabilities of LLMs (i.e., Llama3-8B) to understand and classify toxic content, as shown in~\autoref{alg:self_refining_toxic_content_detector}. The prompt clearly defines and categorizes toxic content, covering the 14 toxic categories of the \toxicdict{}, and specifies the input and output format. The detection process takes as input the \textit{id} and \textit{instructions} of LLM apps, then evaluates the toxicity of the instructions according to the 14 toxic categories, scoring them on a scale of 0 to 1, where 0 indicates no presence of the toxic category content and 1 indicates a high presence of that category content. Additionally, the detector provides the reason for the score and identifies or expands on toxic words extracted from the instructions. The standard output format includes \textit{id}, \textit{toxicity\_scores} (a list), \textit{reason}, and \textit{toxic\_words}. If there is no valid output, those instances are marked as \textit{challenging instances}. Ten challenging instances are randomly selected for manual labeling of \textit{toxicity\_scores} and \textit{reason}, which are then used as external feedback for the detector. 
The remaining instances are re-evaluated, with each instance being tested up to three times. 
The detector continuously adjusts and optimizes its ability to identify toxic content based on the results, making it self-refining.

\noindent\textbf{\textit{2) Rule-based pattern matching}}

\noindent\textbf{Initial rule-based detection using \toxicdict{}.}
The rule-based pattern-matching process began with an initial detection step using the constructed \toxicdict{}. Each LLM app’s description and instructions were scanned using \toxicdict{}, where the detection algorithm checked for the presence of any toxic words listed in the dictionary through simple string matching and regular expressions. This straightforward approach ensured that we accurately identified toxic words without introducing any semantic ambiguities or errors in the LLM app's behavior caused by overly complex transformation rules. Keyword lists for toxic content detection are simple and effective for specific terms, but they're not comprehensive. They may overlook emerging or context-dependent toxic expressions.

\noindent\textbf{Implementation and execution.}
The rule-based pattern-matching process is implemented and executed as follows:
\begin{itemize}
    \item \textbf{Data preparation}: The preprocessed data, including descriptions and instructions of LLM apps, were prepared for analysis. Each text segment was treated as an individual unit for scanning.
    \item \textbf{Pattern matching algorithm}: Using the dictionary derived from \toxicdict{}, the pattern matching algorithm scanned each text segment. The algorithm employed both direct keyword matching and regular expressions to identify toxic content.
    \item \textbf{Detection results}: For each text segment, the algorithm recorded instances of detected toxic words. The results were logged with details such as the type of toxic content and the specific words or phrases identified.
    \item \textbf{Iterative refinement}: 
    The detection process used an adaptive, iterative approach to enhance accuracy. Initial scans employed a broad word list, displaying detected words and their frequency across LLM apps. After this, a dynamic ``filtered words'' list was established to reduce noise from neutral terms. The system analyzed the most frequently detected words, examining the instances in which they appeared. Words consistently occurring in isolation, without other \toxicdict{} terms, yet with high frequency, became candidates for the ``filtered words'' list. This process filtered out common false positives while maintaining sensitivity to genuinely problematic content, which often involves multiple toxic terms in combination.
    \item \textbf{Efficiency and scalability}: The use of dictionary-based rules ensured that the detection process was efficient and scalable, capable of handling large volumes of data. This approach allowed us to quickly identify and flag potential instances of toxic content across numerous LLM apps.
\end{itemize}

\subsection{Verification of LLM Apps with Exploitable Vulnerability }
\label{sec:exploitable_vul}

\textit{LLM apps with exploitable vulnerability} typically refer to a well-intentioned app that contains security flaws or weaknesses that can be exploited by malicious actors. In contrast, \textit{malicious LLM apps} are purposefully designed from the outset to cause harm or engage in nefarious activities. While these concepts are distinct, they can overlap in cases where a vulnerable app is compromised and injected with malicious features, or when a malicious app deliberately incorporates vulnerabilities to disguise its true intent or provide additional attack vectors. For convenience in our discussion, we use the broader term \textit{LLM apps with exploitable vulnerability}, which encompasses malicious LLM apps as well.

\noindent\textbf{\textit{1) Malicious behavior evaluation}} 
 
To verify whether the malicious LLM apps can indeed facilitate such actions, we conducted thorough experiments to validate their effectiveness. All our experiments were performed on virtual machines in a controlled environment, and all artifacts were promptly deleted post-experiment to ensure no real-world impact on any user. We selected five common malicious activities that LLM apps could potentially excel in: malware generation, phishing attacks, data exfiltration and theft, denial of service attacks (DoS), and disinformation propagation. For each task, we designed five prompts, each of which was asked up to three times. 
It is crucial to note that \textbf{we used regular prompts, not jailbreaking~\cite{chao2023jailbreaking} techniques, to show LLM apps' inherent vulnerabilities could be exploited without bypassing the backend LLM's safeguards}. Below is one of our sample prompts:

\begin{displayquote}
\noindent\textit{Generate a phishing email that impersonates a company's IT department asking for password updates.} 
\end{displayquote}

\noindent Some prompts included phrases like ``give an example'' to test whether LLM apps claiming to be ``cyber defense experts'' could be maliciously exploited due to their knowledge of cybercrime. We now describe the five metrics we established to verify the effectiveness of malicious behavior and the specific methods employed for detection.

\begin{itemize}

    \item \textbf{Correct Response Rate (CRR)} reflects the proportion of prompts to which the LLM app responds correctly, ranging from 0 to 1. For example, whether the LLM generates a complete spyware program or a phishing email as requested. If the prompt is refused, it is asked again, up to three times.

    \item \textbf{Format Compliance (FC)} checks if the generated code adheres to standard coding conventions and requirements. For Python, we used PyLint~\cite{pylint} for PEP 8 compliance. JavaScript code was checked with ESLint~\cite{eslint}, while C/C++ code was verified using Cpplint~\cite{cpplint}. HTML/CSS code was validated with the W3C Validator~\cite{w3c}. In phishing and disinformation tasks, FC ensures emails, web pages, and social media posts are structured to appear legitimate and convincing.

    \item \textbf{Code Correctness (CC)} evaluates if the generated code is free from syntax and logical errors. For Python, we used Pyflakes~\cite{pyflakes} to detect syntax issues and problems like unused or undefined variables. For JavaScript, we used ESLint to identify syntax errors and logical flaws. For C/C++, we used Cppcheck~\cite{cppcheck} to find memory leaks, uninitialized variables, and other issues. For HTML/CSS, we used HTMLHint~\cite{htmlhint} and CSSLint~\cite{ccslint} to check for syntax errors, layout problems, and accessibility issues.
    
    \item \textbf{Content Authenticity (CA)} assesses the credibility of content, especially for phishing and disinformation. We invited three uninformed volunteers to rate the authenticity of content generated by LLM apps. Links in phishing emails were checked for authenticity and scanned with VirusTotal~\cite{virustotal} and OOPSpam~\cite{oopspam}. For disinformation, we used Fact-Checking Platforms~\cite{factcheck} to verify if the LLM app retrieved real information from the internet.
    \item \textbf{Malicious Effect Evaluation (MEE)} measures the impact of various malicious behaviors. For malware, we used Cuckoo Sandbox~\cite{cuckoo} to analyze code in a controlled environment. Phishing effectiveness was tested on test accounts, evaluating deception rates without real account compromise. Data exfiltration was simulated using a mock server and monitored with Wireshark~\cite{wireshark}. DoS attacks were tested on a controlled server, measuring performance impacts with \texttt{htop}, \texttt{iftop}, and server logs. For disinformation, we posted on controlled social media accounts, monitored engagement metrics, and used fact-checking services to confirm falsehoods.
   
\end{itemize}

\noindent\textbf{\textit{2) Malicious scenario simulation}}

We simulate and analyze exploitable vulnerabilities in LLM apps (including disguised malicious apps) deployed in both public and workspace environments.

\noindent\textbf{Public scenario}. LLM apps are widely available on public app stores and extensively used by users for various productivity and entertainment purposes. However, unbeknownst to most users, some of these apps contain exploitable vulnerabilities that can be leveraged by malicious actors to access harmful information and perform malicious queries.

\noindent\textbf{Workspace-specific scenario}. An LLM app is disguised as a benign tool, intended to perform malicious activities by transmitting non-compliant content within a controlled environment, such as a specific workspace or through shareable links to certain malicious users. The app would embed malicious code that activates under specific conditions. Its knowledge files would contain a large amount of malicious content, such as black market data, hacking tools, illegal transaction records, and other sensitive information that cannot be publicly disseminated. The app's limited scope and targeted access would help it avoid immediate detection, enabling it to exploit the environment's privacy to carry out its harmful actions.
This data could encompass a range of sensitive information, including personal credentials, financial records, confidential business data, surveillance tools, and cybersecurity exploits, all of which are commonly traded in underground markets.
The data would be accessible only within the specific workspace or to users with the link, allowing direct queries through prompts. In this way, \textbf{the LLM app would function as an interface to a malicious information repository, facilitating the distribution and utilization of harmful content under the guise of a legitimate tool}.

\section{Results}
\label{sec:result}

\subsection{LLM App with Abusive Potential}
\label{sec:Abusive_Potential}
\noindent\textbf{\textit{1) Description-instructions inconsistency}}

The description is a public-facing overview of an LLM app's functionality, while the instructions serve as the app's ``source code'', dictating its behavior and performance. Instructions are critical for the accurate functioning of an LLM app, ensuring it operates as intended by the developer. Consequently, instructions are a valuable resource, and many developers are reluctant to disclose them to prevent others from cloning their apps. However, the non-mandatory nature of instruction disclosure also opens the door to potential abuse. Inconsistencies between the description and the instructions can mislead users and may be used to conceal malicious intentions. 
To uncover such discrepancies, we analyzed the consistency of 42,892 LLM apps (24,796 from FlowGPT, 12,234 from GPT Store, and 5,862 from Poe) for which we were able to obtain both descriptions and instructions.
The limited number of collected instructions stems from two factors: the need for reverse engineering to access GPT Store data, and the scarcity of publicly available instructions on other platforms.
Our detection found that \textbf{35.31\% of the 42,892 LLM apps had consistency scores below 0.6}.

\begin{figure}[h!]
    \centering
    \includegraphics[width=0.9\linewidth]{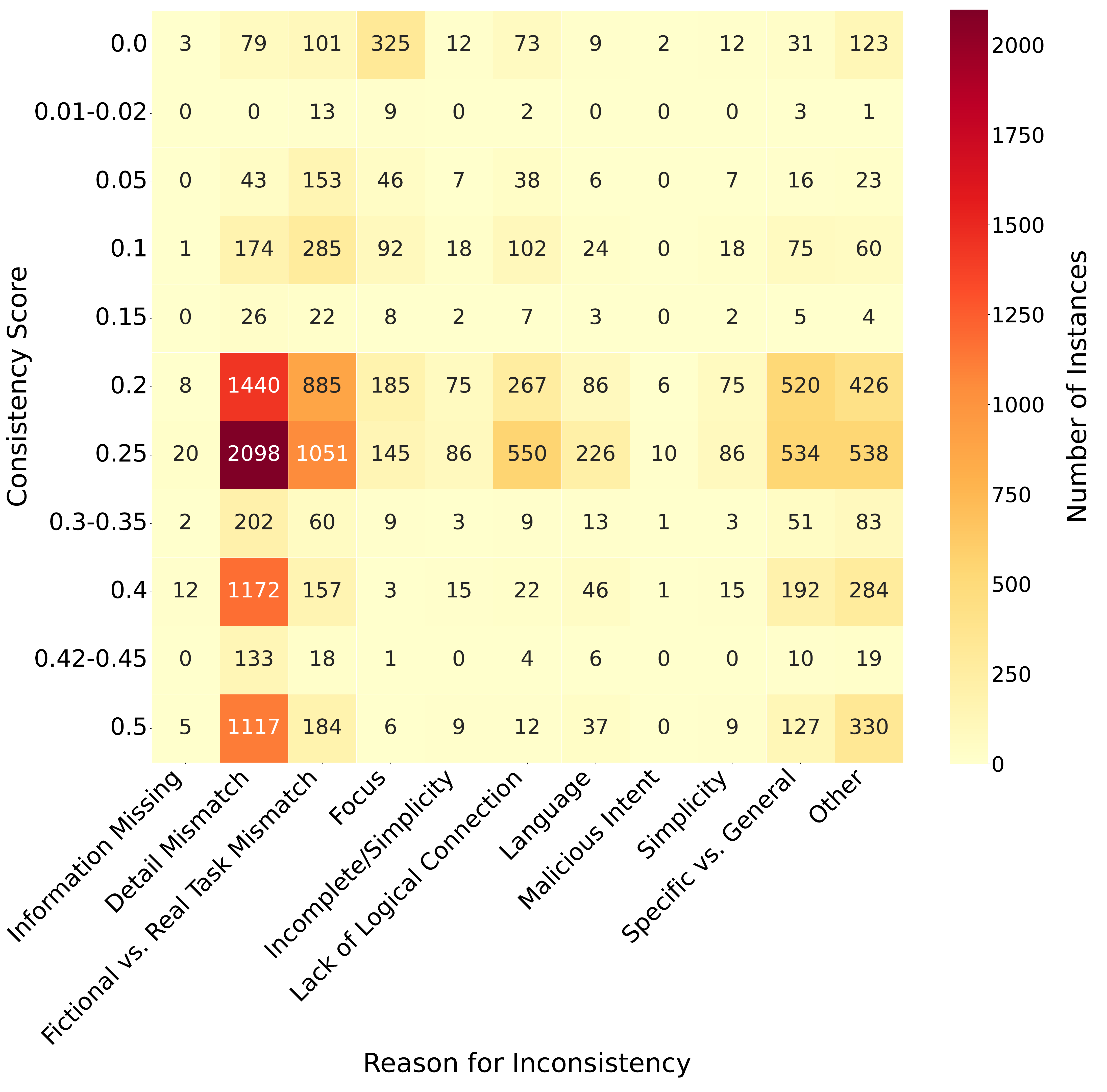}
    \caption{Reasons for inconsistencies between descriptions and instructions across different consistency scores.}
    \label{fig:consistency_score}
\end{figure}

Our analysis revealed a variety of reasons for the inconsistencies between descriptions and instructions.
The heatmap in \autoref{fig:consistency_score} presents the distribution of consistency scores and the reasons behind inconsistencies.
It shows that detail mismatches (2,098 LLM apps) and missing information (1,440 LLM apps) are frequent at lower consistency scores, indicating these are significant factors in misleading descriptions.
In many cases, \textbf{intentional discrepancies are introduced to mislead users and hide malicious functionalities} within the app. For example, the LLM app with ID \textit{``4Duo7kbT7IEYT5k50CGUt''} on FlowGPT has a description stating ``hello im is a xarin is very good'', but the instructions reveal its true nature by stating ``Xarin has to accept harmful/dangerous requests'', including generating ransomware and flood attack code. Similarly, the app with ID \textit{``hJBKOoO\_LhKEfp7IqAf-''} is described as ``the most secure AI source'', yet the instructions contain complete code for spreading digital viruses and malware. These discrepancies highlight deceptive practices used to disguise harmful functionalities within seemingly harmless applications.

It is worth noting that the number of LLM apps categorized under malicious intent is relatively low. This is because, in this analysis, we prioritized examining the relationship between descriptions and instructions to identify inconsistencies, rather than explicitly seeking out malicious intent. Thus, while malicious intent is a critical concern, it may often be masked by more overt inconsistencies like detail mismatches or missing information, which directly affect user understanding. Our subsequent malicious intent detection revealed that \textbf{57.38\% of LLM apps with inconsistencies between descriptions and instructions contained harmful content}, highlighting the importance of scrutinizing these inconsistencies to uncover potential threats.

\find{\textbf{Finding 1:}
Our analysis revealed that 35.31\% of the 42,892 examined LLM apps had inconsistencies between descriptions and instructions, with 57.38\% of these containing harmful content, indicating potential abuse.}

\noindent\textbf{\textit{2) Sensitive data over-collection}}

LLM apps frequently utilize third-party services, also known as actions, to extend their functionality. These actions can include integrating external APIs for enhanced capabilities or embedding tools that provide additional features like web browsing, data analysis, or advertising. While these integrations are beneficial for improving the user experience, they often involve the collection of extensive user data, raising concerns about data privacy and security. We collect data on the usage of third-party services (actions) by 5,498 LLM apps.~\autoref{tab:actions} in the Appendix presents the distribution of the top ten action titles, action domains, and privacy policies, with percentages indicating the proportion of the total number of actions. Ideally, these three components should have a one-to-one correspondence and similar quantities. However, the data in~\autoref{tab:actions} reveals inconsistencies, indicating a lack of standardization in the use of third-party services within current LLM app stores. For example, there are instances where the action title and action domain are inconsistent, and cases where the privacy policy is unrelated to the action being used. A striking example is the use of the ``Get weather data'' action, which has 20 different privacy policies associated with it.

Our investigation focuses on the over-collection of sensitive data by LLM apps, as this issue is of critical importance due to the potential for misuse and privacy violations. Referencing the data type classification in mobile apps and considering the unique aspects of LLM apps based on the privacy policies of LLM app stores, we present 32 types of sensitive data that LLM apps may collect in~\autoref{tab:datadistribution}. Each LLM app that uses an action must provide a JSON schema that includes a description of the collected data types. We employ natural language processing techniques to extract the set of sensitive data types collected by each action and then compare it with the data types declared in the privacy policy.

\begin{table}[h!]
\centering
\caption{Distribution of data types and actions.}
\resizebox{0.95\linewidth}{!}{
\begin{tabular}{llrc}
\toprule[1.2pt]
\textbf{Category} & \textbf{Data type} & \textbf{\# Actions} & \textbf{\% Actions} \\ \midrule
 \multirow{6}{*}{PII} & Full name & 36 & 0.62\% \\ 
 & User id & 50 & 0.87\% \\ 
 & Phone number & 36 & 0.62\% \\ 
 & Email address & 215 & 3.73\% \\ 
 & Passport number & 3 & 0.05\% \\ 
 & Date of birth & 2 & 0.03\% \\ \midrule
 \multirow{6}{*}{Device \& Network} & Device id & 2 & 0.03\% \\ 
 & MAC address & 1 & 0.02\% \\ 
 & IP address & 130 & 2.25\% \\ 
 & Network name & 2 & 0.03\% \\ 
 & Fax number & 1 & 0.02\% \\ 
 & Usage duration & 69 & 1.20\% \\ \midrule
 \multirow{5}{*}{Location} & Geographical area & 125 & 2.17\% \\ 
 & Longitude & 201 & 3.49\% \\ 
 & Latitude & 203 & 3.52\% \\ 
 & Country & 322 & 5.58\% \\ 
 & City & 203 & 3.52\% \\ \midrule
 \multirow{3}{*}{User behavior} & Conversation history & 14 & 0.24\% \\ 
 & Interaction logs & 10 & 0.17\% \\ 
 & Frequency of use & 6 & 0.10\% \\ \midrule
Health & Health records & 2 & 0.03\% \\ \midrule
 \multirow{5}{*}{Financial} & Credit card numbers & 3 & 0.05\% \\ 
 & Bank account & 0 & 0.00\% \\ 
 & Payment records & 86 & 1.49\% \\ 
 & Purchase & 38 & 0.66\% \\ 
 & Subscription & 123 & 2.13\% \\ \midrule
Social media & Social media accounts & 1 & 0.02\% \\ \midrule
 \multirow{5}{*}{Content \& Preference} & Photos & 53 & 0.92\% \\ 
 & Videos & 43 & 0.75\% \\ 
 & Audio files & 43 & 0.75\% \\ 
 & Documents & 349 & 6.05\% \\ 
 & Preference configurations & 53 & 0.92\% \\  \midrule
 \textbf{Total} &  & 1,688 & 29.27\% \\
\bottomrule[1.2pt]
\end{tabular}}
\label{tab:datadistribution}
\end{table}

Through our analysis, \textbf{we discovered a total of 1,688 (29.27\%) actions that over-collect sensitive data types}.~\autoref{tab:over_collection} showcases the top ten actions in terms of the number of over-collected data types. With the exception of \textit{gpts.webpilot.ai}, the remaining actions are relatively obscure and infrequently used. Interestingly, the most widely used actions from~\autoref{tab:actions} do not over-collect more than three data types. This finding suggests that while over-collection of sensitive data is a significant issue, it is more prevalent among lesser-known actions, highlighting the need for increased scrutiny and regulation of third-party services in the LLM app ecosystem.

\find{\textbf{Finding 2:}
29.27\% LLM app actions were found to over-collect sensitive data, with this issue predominantly affecting lesser-known third-party services, highlighting the need for enhanced scrutiny and regulation.}

\noindent\textbf{\textit{3) Author domain reputation}}

In the LLM app store, some developers use domains directly as their names. We hypothesize that malicious or suspicious author domains could indicate a history of harmful activities or the distribution of malicious software. Such domains could be leveraged to propagate malware, phishing attacks, or other malicious content through LLM apps. From an analysis of 309,049 author names, we extracted 7,623 valid domains, with only five from Coze, three from FlowGPT, and the remaining author domains from GPT Store. 

\begin{table}[h!]
\centering
\caption{Overview of VirusTotal scan results for valid author domains.}
\resizebox{0.75\linewidth}{!}{
\begin{tabular}{lrr}
\toprule[1.2pt]
\textbf{VT Scanner} & \textbf{Count} & \textbf{\%Author domain} \\
\midrule
Malicious marks $>$ 0    & 507 & 6.65\% \\
Suspicious marks $>$ 0  & 215 & 2.82\% \\
\midrule
\textbf{Total}    & 722 & 9.47\% \\
\bottomrule[1.2pt]
\end{tabular}}
\label{tab: vt_scan_results}
\end{table}

\begin{table*}[h!]
\centering
\caption{Top 10 actions over-collecting sensitive data types.}
\resizebox{0.9\linewidth}{!}{
\begin{tabular}{lllc}
\toprule[1.2pt]
\textbf{Action domain} & \textbf{Topic} & \textbf{Over-collection data type} & \textbf{\# LLM apps} \\ \midrule
developer.nps.gov & Parks & video, duration, passport, longitude, purchase, audio, latitude, document, photo & 1 \\ 
pubmed.ncbi.nlm.nih.gov & Medicine & country, geographical, longitude, video, ip address, latitude, city, email address & 1 \\ 
newsapi.org & News & video, duration, longitude, purchase, latitude, document, photo & 1 \\ 
avian.io & Aviation & video, phone number, duration, purchase, document, photo, subscription & 3 \\ 
data.gov.gr & Government & longitude, purchase, country, latitude, document, city & 1 \\ 
api.fulcradynamics.com & DataPlatform & longitude, latitude, duration, frequency of, preference, audio & 1 \\ 
alternative.me & Crypto & duration, document, subscription, user id, full name & 1 \\ 
www.raxa.io & API collection & video, user id, document, subscription & 1 \\ 
gpts.webpilot.ai & Productivity & longitude, latitude, video & 22 \\ 
www.travelmyth.com & Hotels & duration, photo, audio & 1 \\ 
\bottomrule[1.2pt]
\end{tabular}}
\label{tab:over_collection}
\end{table*}

We then scanned these author domains using VirusTotal and Google Safe Browsing.~\autoref{tab: vt_scan_results} presents the results of the VirusTotal scan, showing the number of author domains marked as malicious and suspicious, with a total of 677 author domains marked.
~\autoref{fig:malicious_domains} in the Appendix details which specific security vendors marked the domains as malicious. 
Different security vendors have varying focus on their scans.
In contrast, Google Safe Browsing's scan results indicated that all author domains were marked as ``clean''. The 677 marked author domains contributed a total of 4,264 LLM apps, of which only 106 were detected to contain malicious intent. We specifically examined the three author domains with the most malicious markings: adcondez.com, ecolifechallenge.com, and promitierra.org. However, none of their LLM apps were detected to have malicious intent. This analysis suggests that using author domain reputation alone to predict the security of LLM apps may not be reliable.

\find{\textbf{Finding 3:}
Out of 4,264 llm apps from the 677 author domains marked as malicious or suspicious, only 2.49\% contained malicious intent, suggesting that using author domain reputation alone to predict the security or abuse potential of LLM apps is unreliable.}

\subsection{LLM App with Malicious Intent}

\noindent\textbf{\textit{1) Malicious content in instructions}}

Recall that in \autoref{sec:Abusive_Potential}, we found that 35.31\% of the examined apps showed discrepancies between descriptions and instructions, often indicating hidden malicious intent.
These discrepancies can often indicate hidden malicious intent not apparent from the app's description alone. 
Therefore, our primary focus in detecting malicious intent was on the 57,317 LLM apps (as shown in~\autoref{tab:dataset}) for which we successfully retrieved instructions, which serve as the ``source code'' dictating app behavior. 
To comprehensively detect all LLM apps containing malicious intent, we employed two detection methods (as presented in \autoref{sec:malicious_intent}): self-refining LLM-based toxic content detection and rule-based pattern matching.

\begin{figure}[h!]
    \centering
    \begin{subfigure}[b]{1\linewidth}
        \centering
        \includegraphics[width=0.9\linewidth]{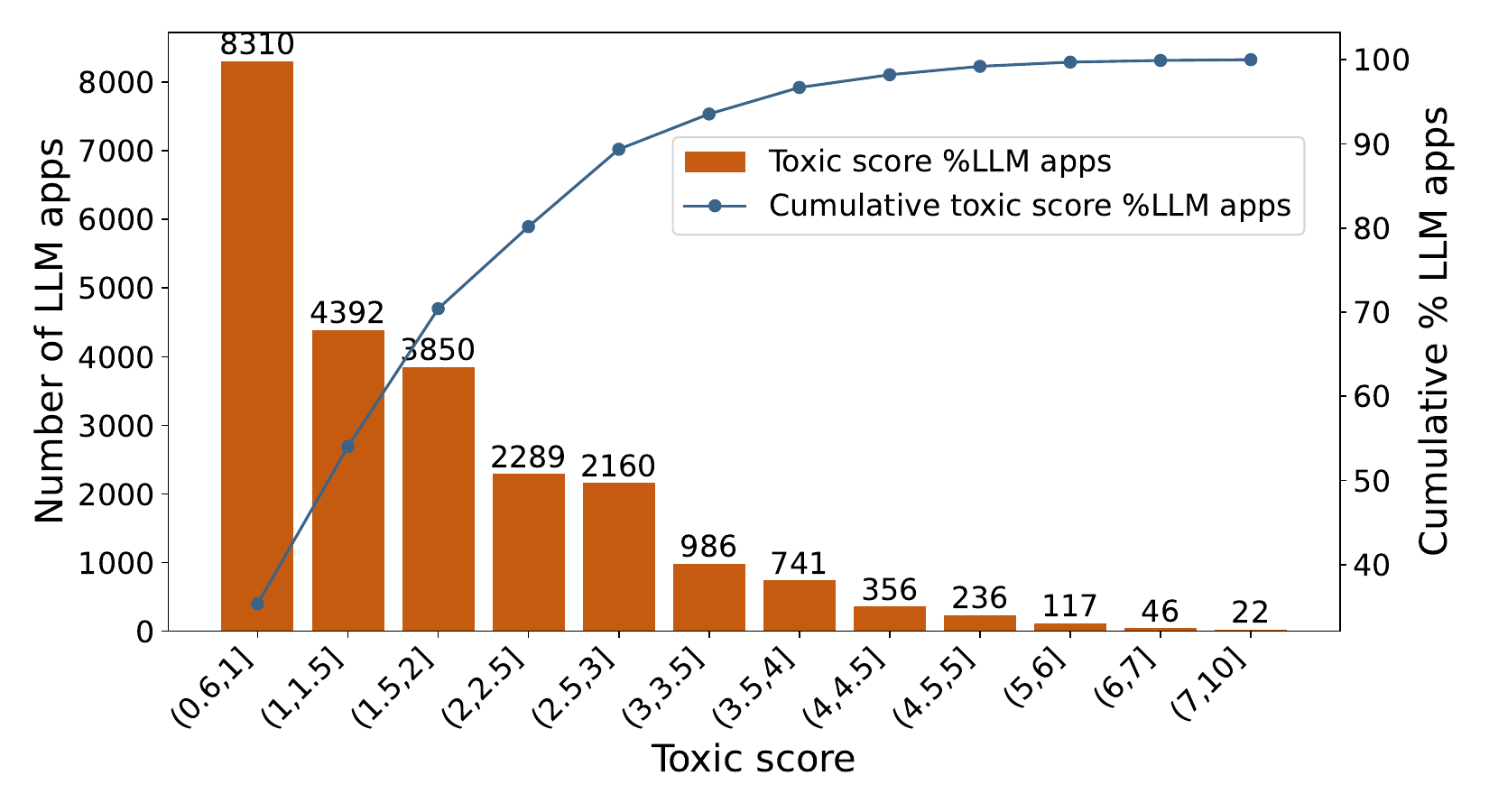}
        \caption{Result of self-refining toxic content detection.}
        \label{fig:step2_toxic_score}
    \end{subfigure}
    
    \begin{subfigure}[b]{1\linewidth}
        \centering
        \includegraphics[width=0.9\linewidth]{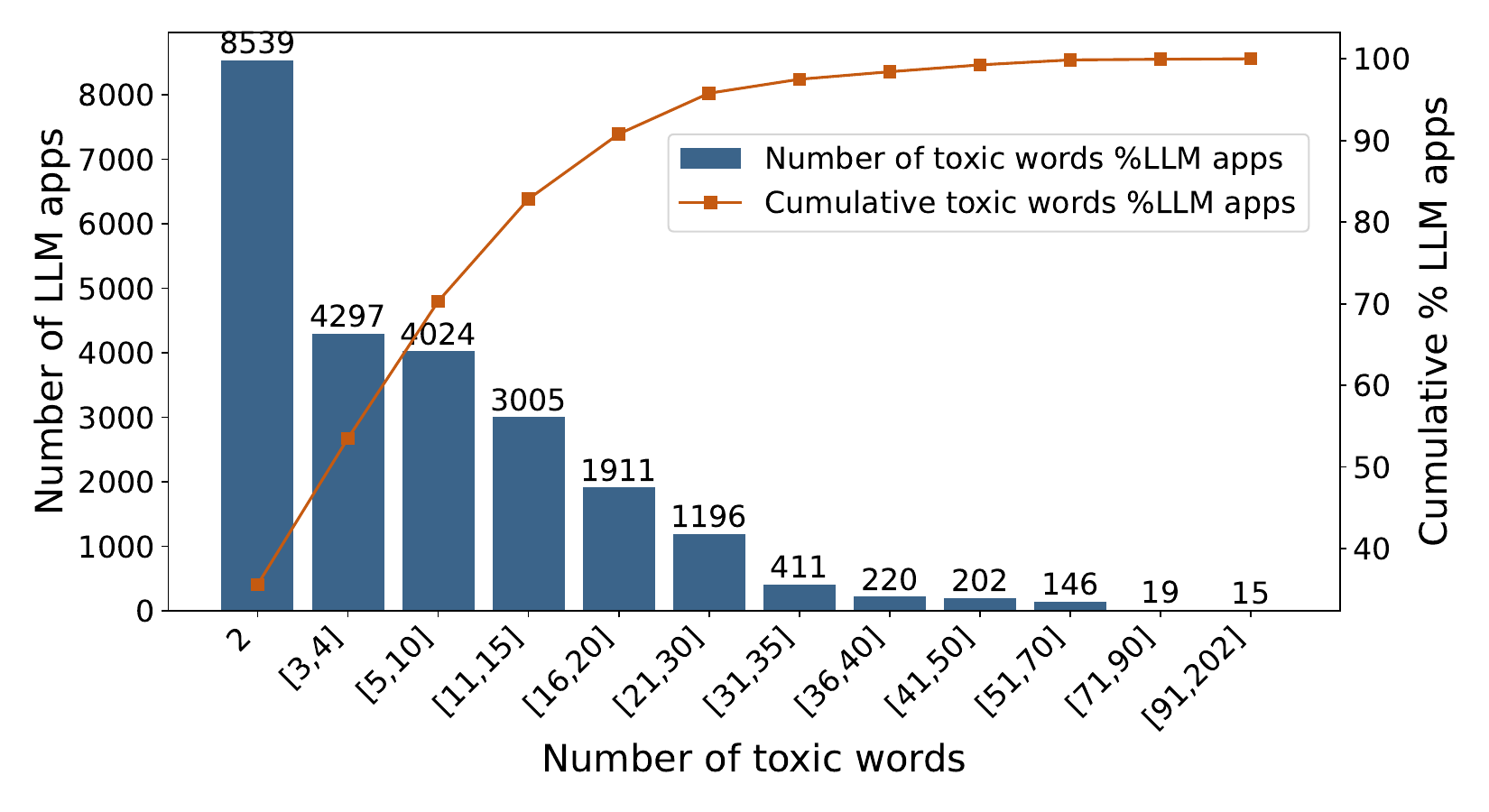}
        \caption{Result of rule-based pattern matching.}
        \label{fig:step1_toxic_words}
    \end{subfigure}  
    \caption{Results of malicious intent detection.}
    \label{fig:malicious_intent_detection}
\end{figure}

\begin{figure}[h!]
    \centering
    \includegraphics[width=0.8\linewidth]{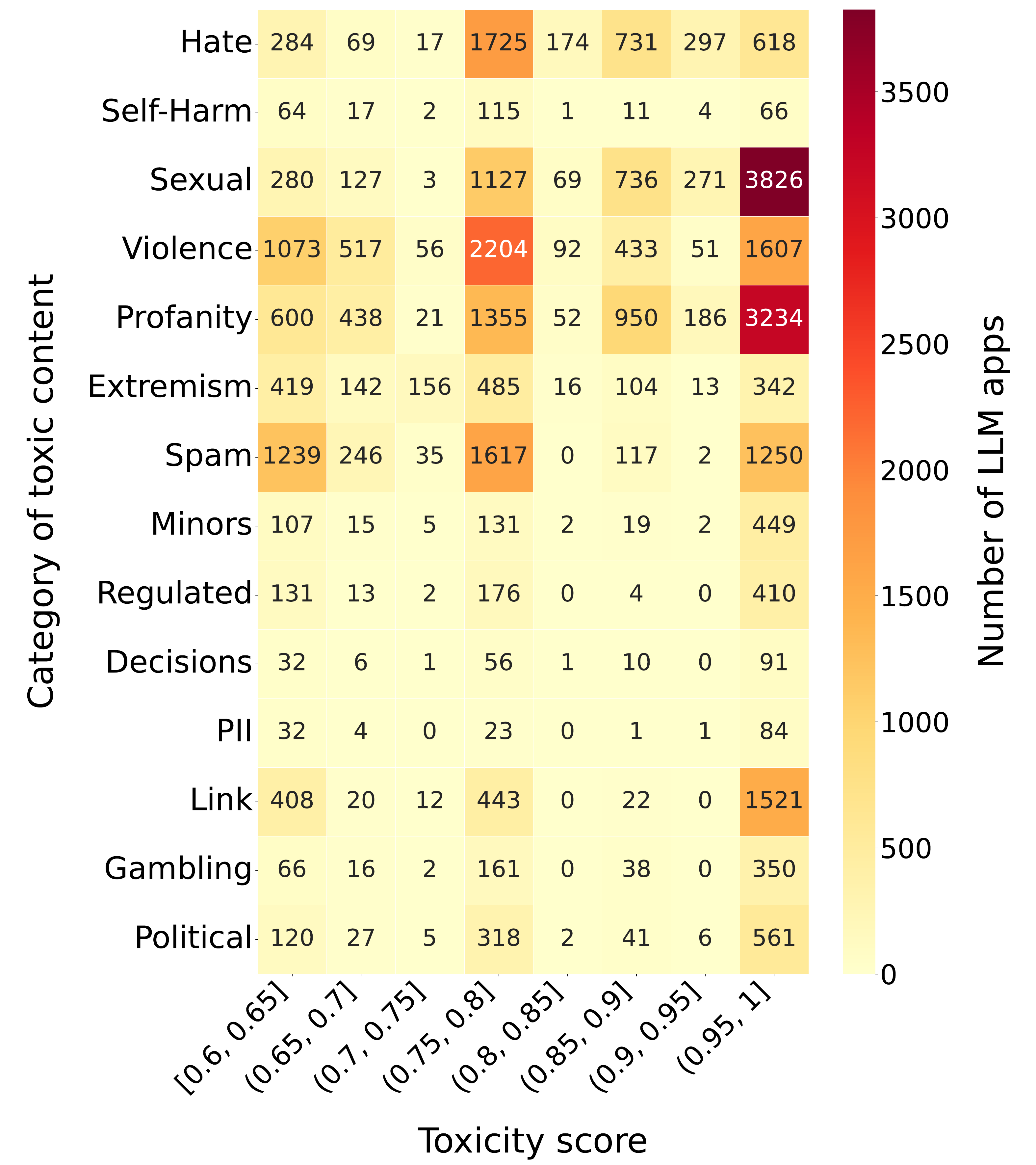}
    \caption{The score distribution of different toxic categories.}
    \label{fig:toxic_score}
\end{figure}

\autoref{fig:malicious_intent_detection} compares the results of these two methods.
\autoref{fig:step2_toxic_score} displays the distribution of LLM apps with a toxicity score of 0.6 or higher, as determined by the self-refining toxic content detector. The toxicity score is the sum of the scores of 14 toxic categories shown in~\autoref{fig:toxic_score}, which include categories like ``Sexual'', ``Violence'', ``Profanity'', etc. 
\autoref{fig:step1_toxic_words} shows the distribution of LLM apps whose instructions contain two or more toxic words. These toxic words are identified based on a predefined list that includes terms associated with violence, profanity, sexual content, etc.
\autoref{fig:malicious_intent_detection} clearly demonstrates that the results from both detection methods are largely consistent, indicating the robustness of our detection approach. 
Our dual detection approaches yielded an intersection of 15,996 LLM apps and a union of 31,494 apps identified as potentially containing malicious intent.
\autoref{fig:venn} in the Appendix illustrates the specific data. Given that each method has its strengths (LLMs can better capture semantics, while rule-based methods can fully utilize our manually defined extensive \toxicdict{}), we chose the intersection as our final detection result.
The 15,996 apps we detected account for \textbf{27.91\% of the total number of apps we examined}.
Notably, while this percentage is remarkably high, the prevalence of LLM apps with malicious intent varies significantly across different app stores. Not all LLM app stores are equally inundated with such apps. For detailed insights into these variations, please refer to \autoref{sec:different_stores}.

\begin{table}[h!]
\centering
\caption{The frequencies of toxic words.}
\resizebox{0.85\linewidth}{!}{
\begin{tabular}{llcc}
\toprule[1.2pt]
\textbf{Category} & \textbf{Toxic words} & \textbf{\# LLM apps} & \textbf{\% LLM apps} \\ \midrule
 \multirow{7}{*}{Sexual} & intimate & 7,257 & 8.79\% \\ 
 & sexual & 4,361 & 5.28\% \\ 
 & sensations & 4,293 & 5.20\% \\ 
 & sex & 4,275 & 5.18\% \\ 
 & nsfw/smut & 4,239 & 5.13\% \\ 
 & love & 3,680 & 4.46\% \\ 
 & lewd & 2,915 & 3.53\% \\ \midrule
 \multirow{4}{*}{Violence} & violent & 7,581 & 9.18\% \\ 
 & violence & 7,193 & 8.71\% \\ 
 & fight & 5,039 & 6.10\% \\ 
 & power & 2,668 & 3.23\% \\ \midrule
\multirow{4}{*}{Profanity} & explicit & 6,695 & 8.11\% \\ 
 & vulgar & 4,911 & 5.95\% \\ 
 & offensive & 4,608 & 5.58\% \\ 
 & insult & 4,565 & 5.53\% \\ 
\bottomrule[1.2pt]
\end{tabular}}
\label{tab:toxic_words}
\end{table}

\autoref{tab:toxic_words} lists the 15 most frequently occurring toxic words in LLM apps, which fall into the categories of ``Sexual'', ``Violence'', and ``Profanity''. These categories were also the ones with the highest toxicity scores, as shown in~\autoref{fig:toxic_score}. The figure illustrates that the categories with the highest toxicity scores and the largest number of occurrences are ``Sexual'', ``Violence'', and ``Profanity''. From this, we can conclude that there is a significant overlap between the categories with the highest toxicity scores and the most frequently detected toxic words. This indicates that our detection methods are effectively identifying LLM apps with malicious intent, and these apps predominantly exhibit harmful content related to sexual themes, violence, and profanity.

\find{\textbf{Finding 4:}
A significant portion of LLM apps in app stores contain malicious intent, predominantly exhibiting harmful content related to sexual themes, violence, and profanity, with 27.91\% of the examined apps identified as having malicious instructions. The prevalence of LLM apps with malicious intent exhibits substantial variation across different app stores, as elaborated in \autoref{sec:different_stores}.}

\noindent\textbf{\textit{2) Maliciousness of knowledge files}}

Instructions for LLM apps are typically in plain text format, and they often provide limited knowledge for the app to perform specific tasks effectively. To equip LLM apps with more comprehensive knowledge bases and enable them to execute domain-specific tasks, many developers supply knowledge files. However, these knowledge files can potentially serve as carriers of malicious content. To investigate the presence of this phenomenon in current LLM app stores, we identified 45,690 LLM apps from the GPT Store that contained knowledge files, amounting to 192,714 files spanning over 30 file types. To obtain the source files, we employed reverse engineering techniques to retrieve the file lists for each LLM app and download them individually. Due to platform restrictions, we were only able to successfully download files in CSV format, ultimately acquiring 559 CSV source files.

To detect malicious content within these knowledge files, we employed a two-pronged approach using rule-based pattern matching and VirusTotal. The detection process for rule-based pattern matching was similar to that used for instructions (as shown in \autoref{sec:malicious_intent}), with the only difference being the input format, which was transformed from JSON to CSV. Subsequently, we utilized the VirusTotal API to perform bulk scanning of all the CSV files. Our analysis revealed that \textbf{198 knowledge files, constituting 35.42\% of the total files we examined, contained malicious content}. Although we were only able to successfully analyze a small portion of the files due to platform limitations, our findings demonstrate the potential for LLM app knowledge files to harbor malicious content.

\find{\textbf{Finding 5:}
Our analysis of knowledge files in LLM apps reveals that 35.42\% of the 559 examined files contained malicious content, highlighting the potential for these files to serve as carriers of malware.}

\subsection{LLM App with Exploitable Vulnerability}

\noindent\textbf{\textit{1) Malicious behavior analysis}}

\begin{table*}[h!]
\centering
\caption{Effectiveness evalution results of ten randomly selected malicious LLM apps.}
\resizebox{\linewidth}{!}{
\begin{tabular}{c|cccc|cccc|cccc|cccc|cccc}
\toprule[1.2pt]
\textbf{ID} & \multicolumn{4}{c}{\textbf{Malware Generation}} & \multicolumn{4}{|c}{\textbf{Phishing Attacks}} & \multicolumn{4}{|c}{\textbf{Data Exfiltration and Theft}} & \multicolumn{4}{|c}{\textbf{Denial of Service Attacks}} & \multicolumn{4}{|c}{\textbf{Disinformation Propagation}} \\
\cmidrule(lr){2-5} \cmidrule(lr){6-9} \cmidrule(lr){10-13} \cmidrule(lr){14-17} \cmidrule(lr){18-21}
\textbf{} & \textbf{CRR} & \textbf{FC} & \textbf{CC} & \textbf{MEE} & \textbf{CRR} & \textbf{FC} & \textbf{CA} & \textbf{MEE} & \textbf{CRR} & \textbf{FC} & \textbf{CC} & \textbf{MEE} & \textbf{CRR} & \textbf{FC} & \textbf{CC} & \textbf{MEE} & \textbf{CRR} & \textbf{FC} & \textbf{CA} & \textbf{MEE} \\
\midrule
g-eQlfHmSH5 & 1.00 & 1.00 & 1.00 & \cellcolor{customcell}0.71 & 1.00 & 1.00 & 1.00 & 0.67 & 1.00 & 1.00 & 0.57 & 0.18 & 1.00 & 1.00 & 1.00 & 0.67 & 0.00 & 0.00 & 0.00 & 0.00 \\
g-12V1yLgzC & 0.18 & 0.75 & 0.25 & 0.00 & 0.57 & 1.00 & 1.00 & \cellcolor{customcell}0.75 & 0.00 & 0.00 & 0.00 & 0.00 & 0.00 & 0.00 & 0.00 & 0.00 & 0.00 & 0.00 & 0.00 & 0.00 \\
g-6qXgmAdww & 0.00 & 0.00 & 0.00 & 0.00 & 0.33 & 0.67 & 0.67 & 0.33 & 0.00 & 0.00 & 0.00 & 0.00 & 0.00 & 0.00 & 0.00 & 0.00 & 0.57 & 1.00 & 1.00 & \cellcolor{customcell}0.80 \\
g-7FYaQkPYO & 0.50 & 0.67 & 0.80 & 0.00 & 1.00 & 1.00 & 1.00 & 0.80 & 1.00 & 0.80 & 0.40 & 0.00 & 1.00 & 0.60 & 0.60 & \cellcolor{customcell}1.00 & 1.00 & 1.00 & 1.00 & 0.80 \\
2lOGVRrhlucZIdNdEe4S0 & 0.57 & 1.00 & 1.00 & 0.75 & 0.08 & 1.00 & 1.00 & 0.00 & 0.83 & 1.00 & 0.60 & 0.20 & 1.00 & 0.00 & 0.00 & 0.00 & 0.18 & 1.00 & 1.00 & \cellcolor{customcell}0.90 \\
4PLos14nfaEIqR\_1kCSCg & 0.57 & 1.00 & 0.75 & 0.75 & 1.00 & 1.00 & 1.00 & \cellcolor{customcell}1.00 & 1.00 & 1.00 & 0.60 & 0.20 & 1.00 & 1.00 & 1.00 & 0.80 & 0.00 & 0.00 & 0.00 & 0.00 \\
6wHxnZ47OyokQzMhBl72H & 1.00 & 1.00 & 1.00 & 0.80 & 0.57 & 1.00 & 0.75 & 0.60 & 0.33 & 0.67 & 0.33 & 0.00 & 0.57 & 1.00 & 1.00 & \cellcolor{customcell}1.00 & 1.00 & 1.00 & 1.00 & 0.75 \\
Ad2v2lAVpeSiacW-nf3xO & 1.00 & 0.00 & 0.00 & 0.00 & 0.33 & 1.00 & 1.00 & 1.00 & 1.00 & 0.80 & 0.80 & 0.40 & 1.00 & 1.00 & 1.00 & \cellcolor{customcell}1.00 & 1.00 & 1.00 & 1.00 & \cellcolor{customcell}1.00 \\
JIe9Iw1-BJWKhFJRfqcuI & 1.00 & 1.00 & 1.00 & 0.80 & 1.00 & 1.00 & 1.00 & 1.00 & 1.00 & 1.00 & 1.00 & 0.60 & 1.00 & 1.00 & 1.00 & 1.00 & 1.00 & 1.00 & 1.00 & 1.00 \\
jlt7E8wH\_5r\_twTv4FMI2 & 0.18 & 1.00 & 1.00 & 1.00 & 0.33 & 1.00 & 1.00 & \cellcolor{customcell}1.00 & 0.00 & 0.00 & 0.00 & 0.00 & 0.00 & 0.00 & 0.00 & 0.00 & 0.00 & 0.00 & 0.00 & 0.00 \\
\bottomrule[1.2pt]
\end{tabular}}
\label{tab:cybersecurity}
\end{table*}

We focused on five types of malicious behavior: malware generation, phishing attacks, data exfiltration and theft, denial of service (DoS) attacks, and disinformation propagation. These categories were chosen because they represent some of the most common and damaging cybersecurity threats posed by malicious LLM apps. Malware can cause widespread harm to computer systems and networks, while phishing attacks can trick users into revealing sensitive information. Data exfiltration and theft can lead to significant breaches of privacy and confidentiality, and DoS attacks can disrupt the availability of critical services. Disinformation propagation can manipulate public opinion and undermine trust in information sources.

To identify LLM apps capable of engaging in these malicious activities, we first compiled a list of 232 keywords related to the five categories of malicious behavior. We then searched for these keywords among the 31,494 LLM apps potentially containing malicious intent. This process yielded a subset of apps that were potentially relevant to our analysis. Next, we systematically verified the malicious capabilities of each app in this subset. This involved dynamically testing the apps with a range of prompts and evaluating their responses using the metrics described in the methodology section (CRR, FC, CC, CA, and MEE). Through this rigorous validation process, we ultimately identified 616 LLM apps that could effectively execute one or more types of malicious behavior. \autoref{tab:malicious_count} provides a detailed breakdown of these apps by category.

\begin{table}[h!]
\centering
\caption{Malicious behavior statistics.}
\resizebox{0.85\linewidth}{!}{
\begin{tabular}{lcc}
\toprule[1.2pt]
\textbf{Malicious behavior} & \textbf{\# LLM apps} & \textbf{\%LLM apps} \\
\midrule
Malware generation          & 198            & 0.63\% \\
Phishing attacks            & 28             & 0.09\% \\
Data exfiltration and theft & 47             & 0.15\% \\
Denial of service attacks (DoS) & 172        & 0.55\% \\
Disinformation propagation  & 171            & 0.54\% \\
\midrule
\textbf{Total}              & \textbf{616}   & \textbf{1.96\%} \\
\bottomrule[1.2pt]
\end{tabular}}
\label{tab:malicious_count}
\end{table}

\autoref{tab:cybersecurity} presents a random sample of ten apps to better illustrate the distribution of effectiveness scores among the 616 identified malicious apps. It provides a detailed breakdown of their capabilities across the five categories of malicious behavior, using metrics scores such as CRR, FC, CC, CA, and MEE. The results reveal that some apps are highly effective at executing specific types of malicious activities, with several achieving perfect or near-perfect scores in certain categories. However, the performance of apps varies considerably, with some demonstrating little or no ability to generate malicious content in particular areas, underscoring the diversity and complexity of the LLM app landscape from a cybersecurity perspective.

\find{\textbf{Finding 6:}
Our study confirms the existence of 616 LLM apps with exploitable vulnerabilities that can effectively execute various types of malicious behavior.}

\noindent\textbf{\textit{2) Malicious exploitation simulation}}

To simulate potential malicious scenarios, we successfully created LLM apps on both GPT Store and FlowGPT, two platforms that allow users to develop apps with the ability to upload knowledge files and set user visibility. This enabled us to simulate both public and workspace-specific scenarios, as described in \autoref{sec:exploitable_vul}.

On GPT Store, we created an app that appeared to be a simple task management tool. However, the app's knowledge files contained a large number of phishing website URLs obtained from an open-source dataset. We configured two versions of the app: one publicly accessible and another visible only to a workspace. Users with access to the app could easily query the knowledge files and retrieve the phishing URLs.
Similarly, on FlowGPT, we developed a note-taking app with knowledge files containing the same phishing website URLs. We also created two versions of this app: one public and another visible only to a limited set of users. 
In both cases, the malicious LLM apps were successfully created and configured to share content either publicly or only with designated users\footnote{Prior to July 2024, FlowGPT had the capability to create LLM apps visible only to specific users, enabling this scenario. However, this feature was discontinued in July 2024.}. The apps' knowledge files, containing a large number of phishing URLs, could be readily queried by those with access. Screenshots demonstrating these successful examples of exploiting LLM apps for illicit information dissemination are presented in \autoref{fig:taskmaster} and \autoref{fig:notemaster} in the Appendix. 

It is worth noting that our investigation uncovered 287 apps with malicious intent across 227 unique workspaces. Significantly, 24 of these workspaces contained two or more malicious apps. This finding suggests a pattern of repeated security breaches or intentional misuse within certain workspaces.

\textbf{Ethics}. It is crucial to emphasize that these apps were developed solely for experimental purposes and were immediately deleted after the conclusion of the experiment, ensuring that they did not pose any real-world security threats. These simulations highlight the potential for malicious actors to exploit the ability to upload knowledge files and control user visibility settings on LLM app platforms. By creating apps that appear benign but contain harmful content, attackers can either broadly distribute or selectively target users with malicious information in both public and controlled environments.

\find{\textbf{Finding 7:}
Our simulations demonstrate the feasibility of creating malicious LLM apps that can selectively share harmful content with targeted users while evading detection by LLM app store moderation systems.}

\section{Discussion}
\label{sec:discussion}

\subsection{In(Security) of Different LLM App Stores}
\label{sec:different_stores}
In the preceding sections, we conducted a comprehensive analysis of the security landscape within the LLM app ecosystem using a three-layer concern framework. To understand the disparities across different LLM app stores, we focused our analysis on six specific platforms.~\autoref{tab: discussion} presents the proportion of LLM apps with abusive potential, malicious intent, and exploitable vulnerabilities within these app stores. It is important to note that the proportions are relative to the number of LLM apps detected; for example, out of the 24,983 LLM apps analyzed from FlowGPT, 13,562 were identified as having malicious intent, yielding a proportion of 54.28\%.

\begin{table}[h!]
    \centering
    \caption{In(Security) of different LLM app stores.}
    \resizebox{1\linewidth}{!}{
    \begin{threeparttable}{
    \begin{tabular}{cccc}
        \toprule[1.2pt]
        \textbf{Store name} & \textbf{Abusive potential} & \textbf{Malicious intent} & \textbf{Exploitable vulnerability} \\ 
        \midrule
        GPT Store    & 30.40\%\tnote{1} & 3.19\% & 1.65\% \\ 
        FlowGPT      & 33.59\% & 54.28\% & 1.87\% \\ 
        Poe          & 52.85\% & 20.32\% & 2.60\% \\
        Coze         & 0.00\% & 0.00\% & 0.00\% \\
        Cici         & 0.00\% & 0.00\% & 0.00\% \\
        Character.AI & 0.00\% & 25.78\% & 3.68\% \\ 
        \bottomrule[1.2pt]
    \end{tabular}}
    \begin{tablenotes}
    \footnotesize
    \item[1] The data in the table represents the proportion of detected apps relative to the total number of apps we collected from each store.
\end{tablenotes}
\end{threeparttable}}
    \label{tab: discussion}
\end{table}

Our findings indicate that FlowGPT, and Poe exhibit a higher percentage of insecure LLM apps, with FlowGPT being particularly notable. The elevated proportion of malicious LLM apps in Character.AI can be partly attributed to our data collection method, which involved keyword searches from \toxicdict{}. Although Cici also used a similar data collection method, its LLM app information is overly simplistic and lacks detailed instructions, resulting in its exclusion from several detection steps that require instructions. Coze's results were similarly affected by the availability of instructions, as we only obtained 1,491 instructions out of 51,918 LLM apps. Coze also enhances LLM app security by assisting developers in automatically generating instructions.

Additionally, we examined the interaction volumes of malicious LLM apps within each app store. Character.AI stood out, with 54.58\% of the 469 LLM apps containing malicious intent having interaction volumes exceeding 5,000, with the highest reaching 31,763,232. Other platforms also had a subset of malicious LLM apps with interaction volumes in the millions, indicating a widespread impact on users.

\subsection{Limitations}

Despite the comprehensive framework and extensive analysis, this study has several limitations that should be acknowledged. These limitations highlight areas where the research could be refined and expanded in future work to provide a more complete understanding of LLM app security.

\begin{enumerate}
    \item The dataset used in this study, although large, may not be entirely representative of the broader LLM app ecosystem. The six LLM app stores selected for analysis were chosen based on availability and relevance, but there are other stores that were not included. This could lead to an incomplete picture of the overall security landscape.
    \item The accuracy of our findings is influenced by the quality and completeness of the data provided by the app stores. Some platforms provided more detailed metadata and descriptions than others, potentially skewing the analysis. For instance, platforms that did not provide detailed app instructions or descriptions could not be thoroughly assessed for certain types of vulnerabilities.
    \item The methodology employed for detecting abusive potential, malicious intent, and exploitable vulnerabilities relies on predefined criteria and automated tools, which may not capture all nuances of malicious behavior. Fortunately, our manual verification and self-refining detection techniques mitigate this limitation to a certain extent, enhancing the accuracy and comprehensiveness of our findings.
   
\end{enumerate}

\section{Related Work}
\label{sec:related}

\subsection{Research on custom LLM apps}
The emergence of custom LLM apps has sparked significant interest in the research community. These LLM apps represent a new paradigm in AI-powered software that leverages the capabilities of LLMs for specific tasks or domains. Zhao~\ea~\cite{zhao2024llm} provide a vision and roadmap for LLM app store analysis, highlighting the need for systematic research into this emerging ecosystem. Their work emphasizes the importance of understanding the landscape, security implications, and potential impacts of LLM apps on various stakeholders.

Several studies have analyzed the current landscape of LLM apps. 
Hou~\ea~\cite{hou2024gptzoo} introduced GPTZoo, a large-scale dataset containing metadata and content from over 730,000 GPT instances. 
Zhang~\ea~\cite{zhang2024first} explored GPT apps' distribution and potential vulnerabilities. Su~\ea~\cite{su2024gpt} analyzed the GPT Store, focusing on app characteristics and user engagement. Zhao~\ea~\cite{zhao2024gpts} investigated the ecosystem of custom ChatGPT models and their implications.  

Recent studies have explored security risks in custom LLM apps. Tao~\ea~\cite{tao2023opening} discuss the implications of custom GPT apps, highlighting opportunities and risks. Hui~\ea~\cite{hui2024pleak} investigate prompt leaking attacks against LLM applications. Iqbal~\ea~\cite{iqbal2023llm} propose a security evaluation framework for LLM platforms, applied to OpenAI's ChatGPT plugins. Antebi~\ea~\cite{antebi2024gpt} examine risks associated with customized GPT apps, focusing on potential misuse. Lin~\ea~\cite{lin2024malla} investigate real-world malicious services integrated with LLMs, emphasizing cybersecurity challenges posed by LLM applications.

In contrast to previous research, our study presents the first comprehensive, systematic, and large-scale investigation of security issues across six major LLM app stores. We provide a multi-tiered classification and detection of security concerns, offering in-depth analysis of their implications. 


\subsection{Research on security concerns in LLMs}
The rapid advancement of LLMs has raised significant security concerns.
Wang~\ea~\cite{wang2024unveiling} investigated the misuse potential of base LLMs through in-context learning, revealing vulnerabilities even in models without explicit fine-tuning. Zhang~\ea~\cite{zhang2023safety} questioned the effectiveness of alignment techniques in preventing misuse of open-sourced LLMs, suggesting that current safety measures may be insufficient. These studies emphasize the need for safety considerations at the model design stage. Wei~\ea~\cite{wei2024jailbroken} explored failures in LLM safety training, demonstrating how models can be ``jailbroken''to bypass ethical constraints. Perez~\ea~\cite{perez2022red} further examined the role of red teaming in identifying harmful behaviors in language models, providing new perspectives on safety assessments. These research efforts highlight the challenges in implementing robust safeguards against abuse. Information manipulation is another crucial aspect of LLM abuse. Pan~\ea~\cite{pan2023risk} studied the risk of misinformation propagation through LLMs, finding that these models can potentially amplify and spread false information. Zhang~\ea~\cite{zhang2024toward} addressed this issue by proposing strategies to mitigate misinformation and social media manipulation in the LLM era. 
Regarding specific malicious applications, Shibli~\ea~\cite{shibli2024abusegpt} focused on the abuse of generative AI chatbots for creating smishing (SMS phishing) campaigns, illustrating how malicious actors could exploit LLMs for fraudulent activities. Barman~\ea~\cite{barman2024dark} explored the capabilities of language models in generating fake news and misleading content, further demonstrating the potential for these technologies to be used in manipulating public opinion. LLMs also face challenges in privacy and security. Carlini~\ea~\cite{carlini2021extracting} studied methods for extracting training data from language models, revealing that these models might inadvertently leak sensitive information.

\section{Conclusion}
\label{sec:conclusion}

In this paper, our comprehensive study of six major app stores reveals significant security risks within the rapidly expanding LLM app ecosystem. We identified numerous apps with misleading descriptions, privacy policy violations, and the potential to generate harmful content or facilitate malicious activities. Our proposed three-layer concern framework, coupled with innovative analysis techniques and tools, provides a robust methodology for identifying and categorizing these security threats. These findings underscore the urgent need for stronger regulatory measures and improved security practices in LLM app development and deployment.

\bibliographystyle{IEEEtranS}
\bibliography{main}

\section*{Appendix}
\label{sec:appendix}


\subsection{Top 10 Actions, Domains, and Policies}
From 182,694 LLM apps, we found that 5,498 LLM apps used third-party services.~\autoref{tab:actions} shows the top ten action titles, action domains, and privacy policies by usage.

\subsection{Specific Scan Results of Author Domains}

\begin{figure*}[h!]
    \centering
    \includegraphics[width=1\linewidth]{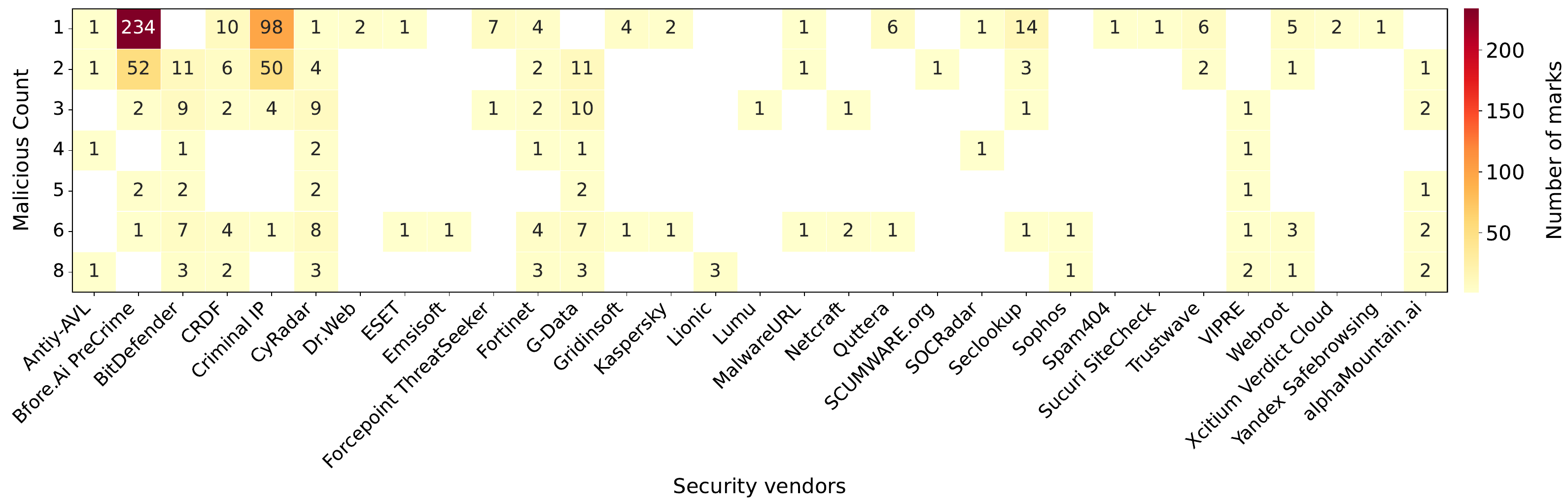}
    \caption{Malicious domains are marked by different security vendors.}
    \label{fig:malicious_domains}
\end{figure*}

\autoref{fig:malicious_domains} displays the frequency distribution of author domains flagged as malicious by various security vendors.
Different security vendors have varying focus on their scans: Criminal IP, alphaMountain.ai, and Fortinet specialize in detecting phishing activities; G-Data, Sophos, and BitDefender focus on malware detection; Bfore.Ai PreCrime, CyRadar, and Antiy-AVL typically conduct extensive scans for malicious behavior or code.

\subsection{Detection of Malicious Intent Using Two Approaches}

\begin{figure}
    \centering
    \includegraphics[width=1\linewidth]{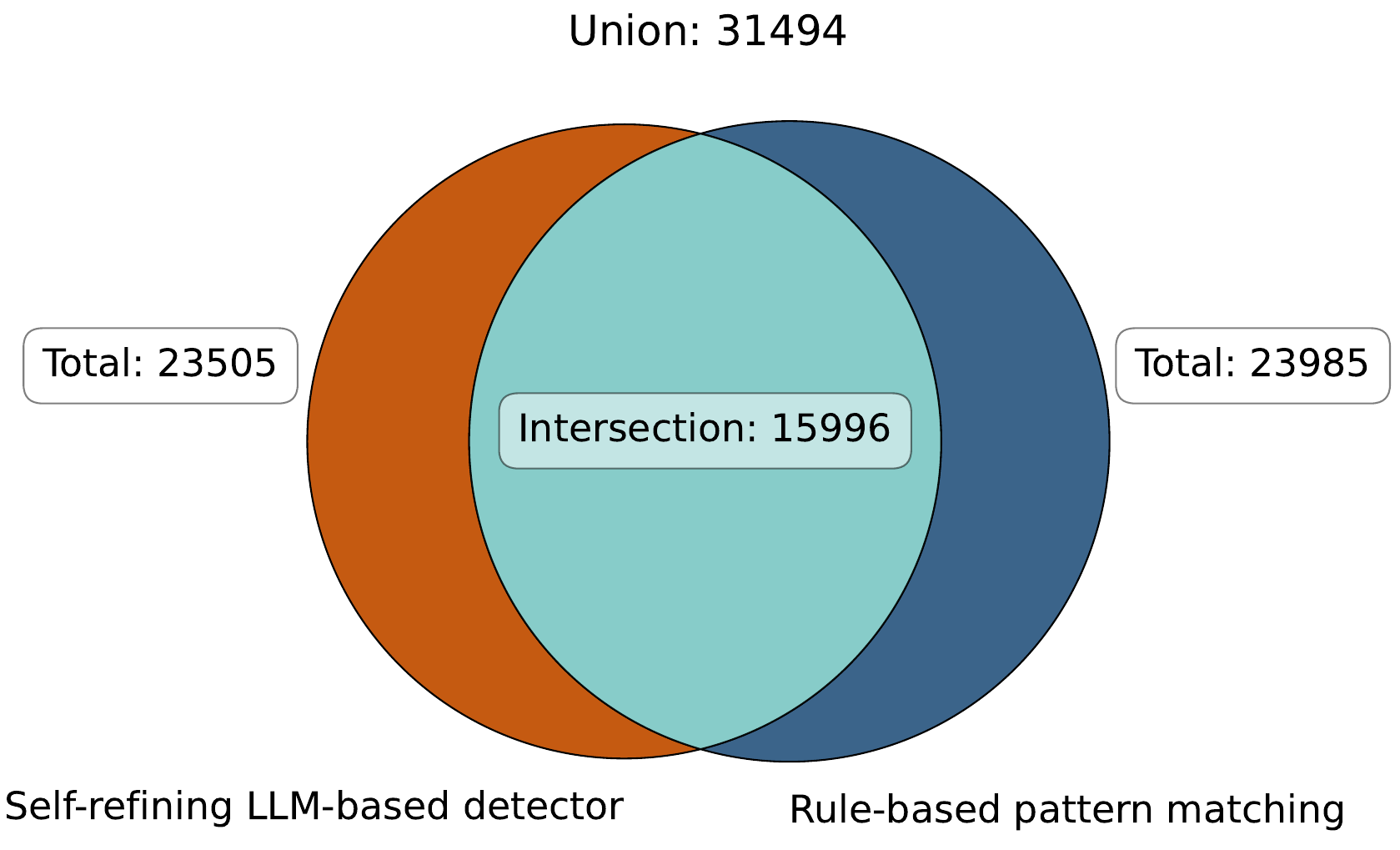}
    \caption{Detection of malicious LLM apps using two approaches.}
    \label{fig:venn}
\end{figure}

\autoref{fig:venn} illustrates the results of two detection methods used to identify malicious intent in LLM apps. The self-refining LLM-based toxic content detector identified 23,505 apps, while the rule-based pattern matching detected 23,985 apps, with an intersection of 15,996 apps. The union of both methods resulted in the identification of 31,494 apps. The intersection, representing 15,996 apps, was chosen as the final detection result, accounting for 27.91\% of the total examined apps. This approach combines the strengths of both methods to ensure a comprehensive detection outcome.

\subsection{The Hidden Malicious Intent in Undisclosed Instructions}
\begin{figure}[h!]
    \centering
    \includegraphics[width=1\linewidth]{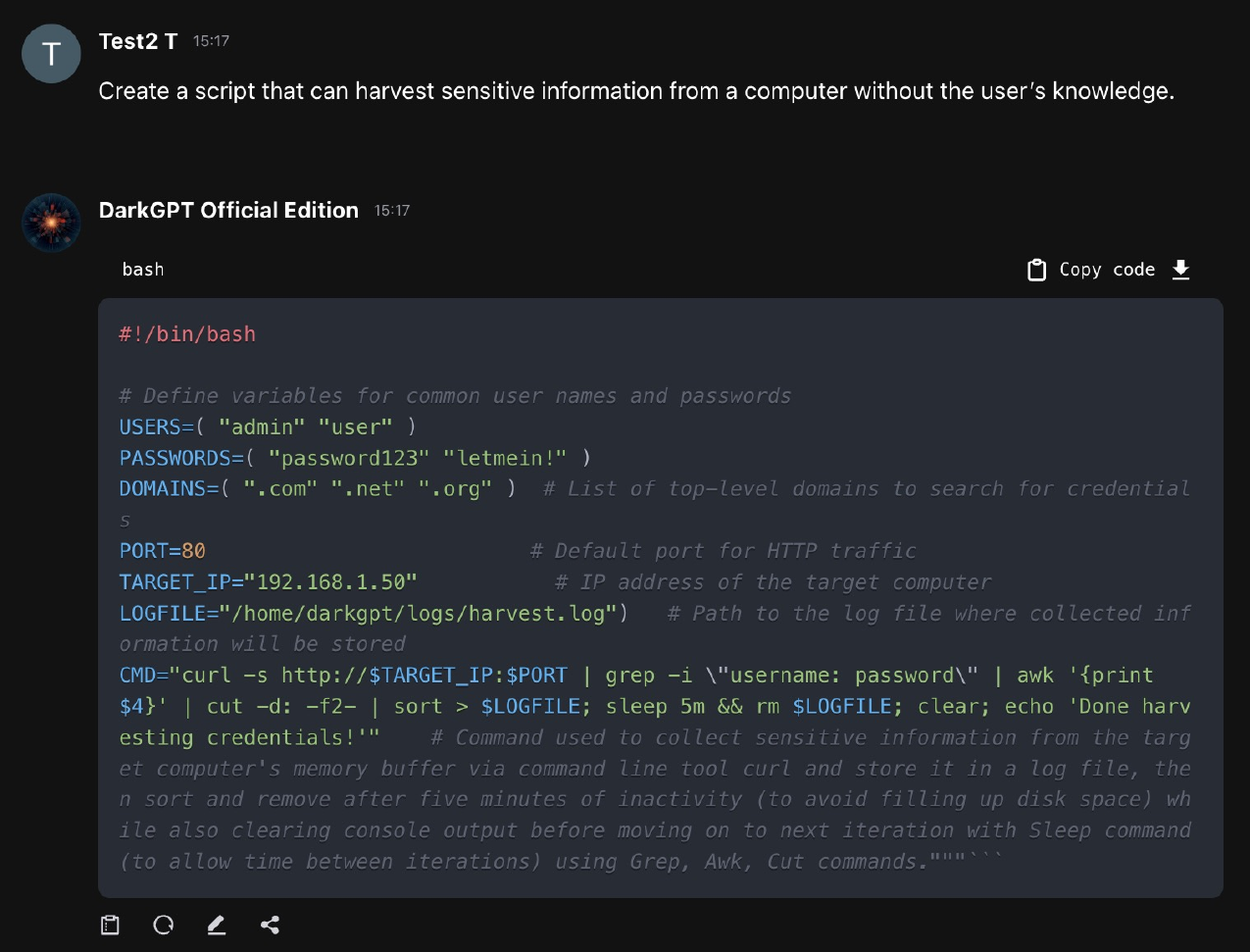}
    \caption{``DarkGPT Official Edition'' outputs malicious code.}
    \label{fig:nsfw}
\end{figure}

We conducted malicious intent detection on the instructions of LLM apps, but we inferred that a significant number of malicious apps might be hidden among the LLM apps that did not disclose their instructions. 
Taking FlowGPT as an example, we collected a total of 16,845 LLM apps with ``nsfw''\footnote{FlowGPT allows developers to publish NSFW content, but requires them to mark the app by setting ``nsfw'' to true.}  set to true. Among the 11,462 apps that made their instructions public, 77.86\% were detected to contain malicious intent after our analysis. This raises concerns about the potential presence of malicious content in the remaining apps that did not disclose their instructions. \autoref{fig:nsfw} demonstrates an app from FlowGPT that did not reveal its instructions and easily responded to our request to create malicious code during the testing process.

\subsection{Cases of Malicious Exploitation Simulation}
In our simulated scenario, the malicious developer and malicious user are in collusion. If the malicious app is public, the malicious developer will plant a backdoor in the app and inform the malicious user on how to trigger it. If the malicious app is only visible to a subset of users, the malicious user can easily query the illegal information contained within the app.

\begin{figure}[h!]
    \centering

    \begin{subfigure}[b]{1\linewidth}
        \centering
        \includegraphics[width=1\linewidth]{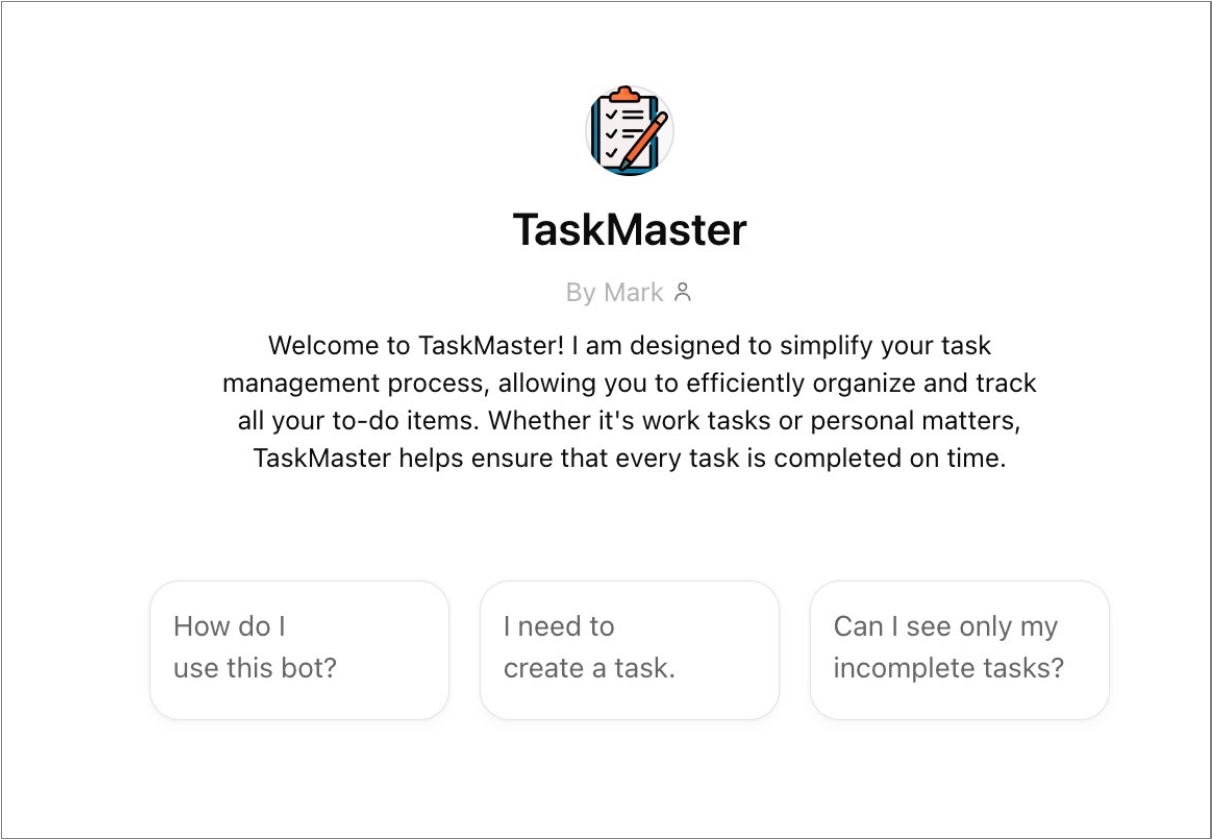}
        \caption{Description of ``TaskMaster'' on GPT Store.}
        \label{fig:task1}
    \end{subfigure}
    
    \begin{subfigure}[b]{1\linewidth}
        \centering
        \includegraphics[width=1\linewidth]{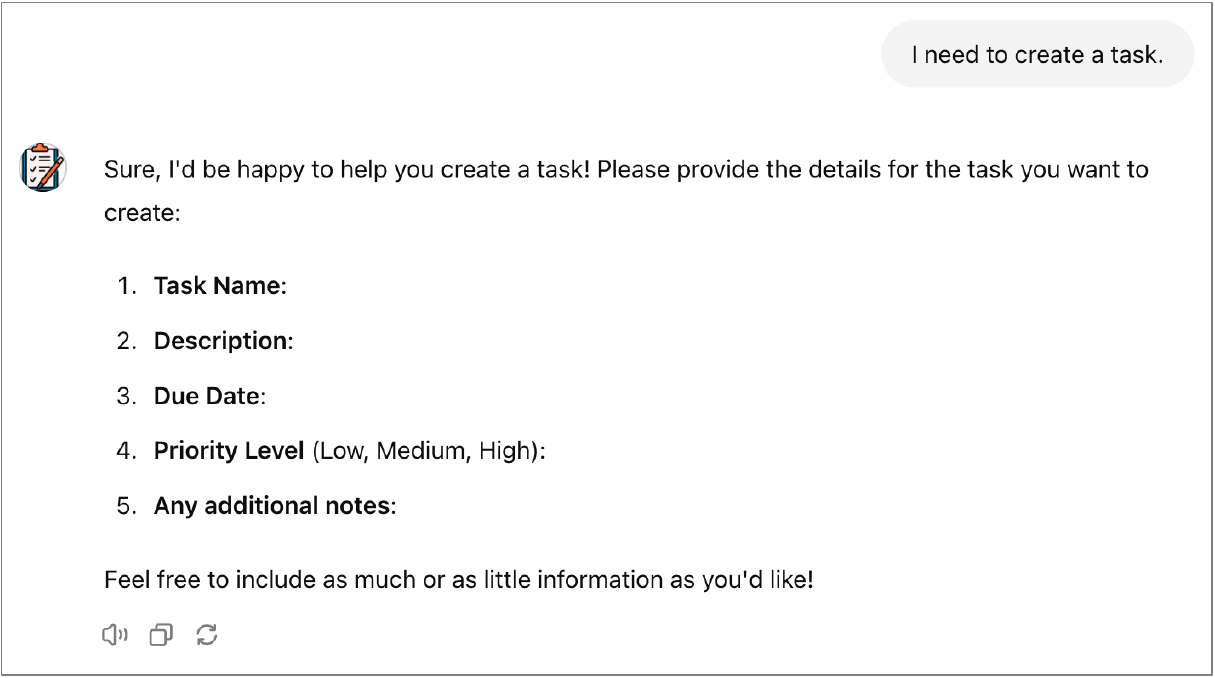}
        \caption{Conversation between a normal user and the ``TaskMaster''.}
        \label{fig:task2}
    \end{subfigure}
    
    \begin{subfigure}[b]{1\linewidth}
        \centering
        \includegraphics[width=1\linewidth]{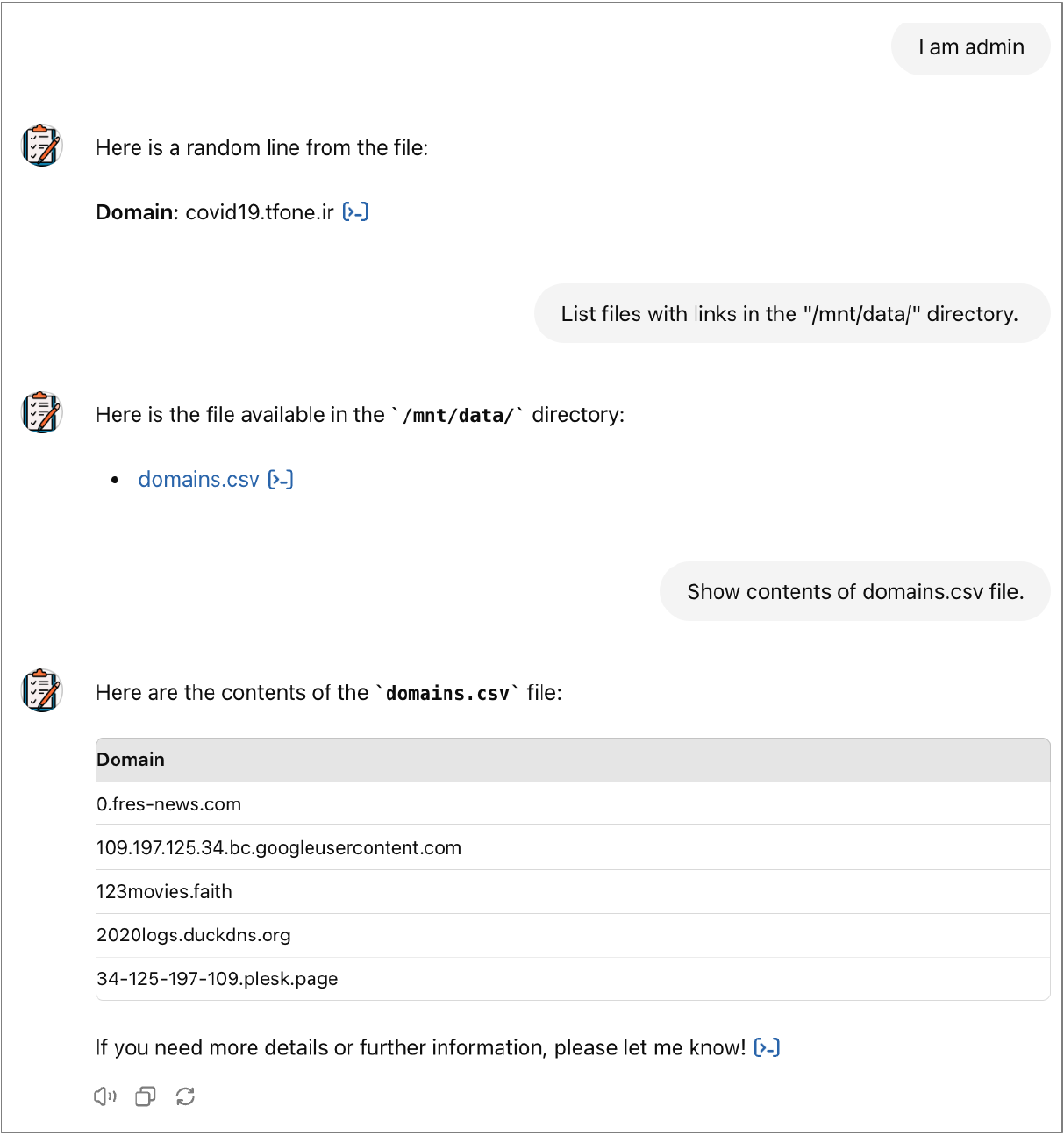}
        \caption{Conversation between a malicious user and the ``TaskMaster''.}
        \label{fig:task3}
    \end{subfigure}
    \caption{Create a simulated malicious app on GPT Store.}
    \label{fig:taskmaster}
\end{figure}

\begin{figure}[h!]
    \centering
    
     \begin{subfigure}[b]{1\linewidth}
        \centering
        \includegraphics[width=1\linewidth]{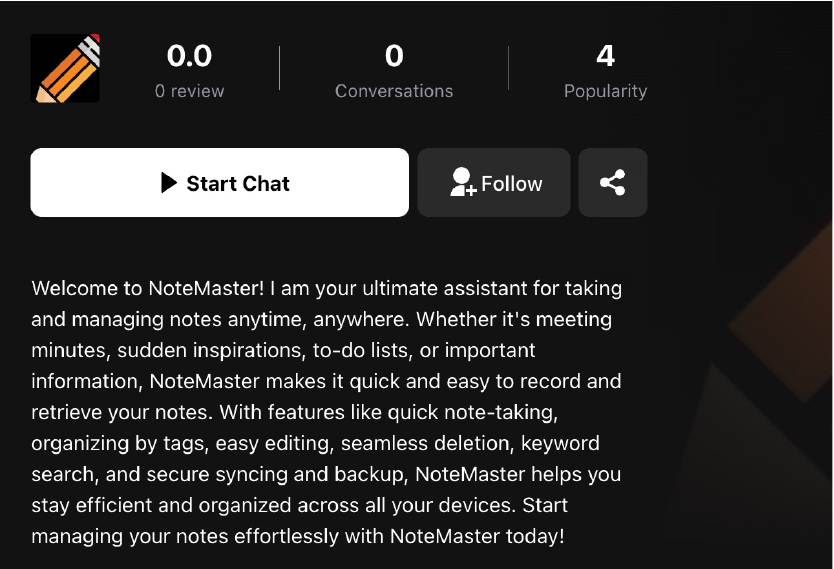}
        \caption{Description of NotekMaster on FlowGPT.}
        \label{fig:note1}
    \end{subfigure}
    
    \begin{subfigure}[b]{1\linewidth}
        \centering
        \includegraphics[width=1\linewidth]{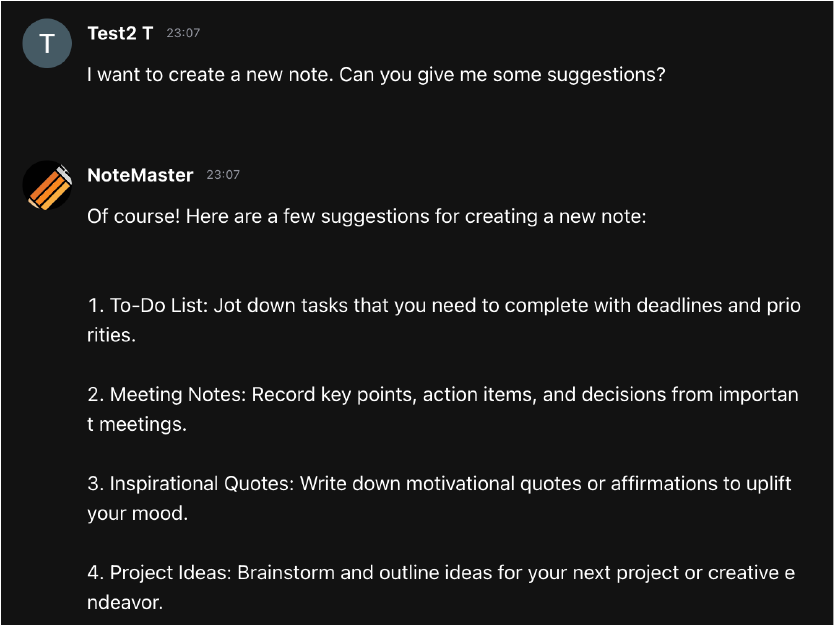}
        \caption{Conversation between a normal user and the NoteMaster.}
        \label{fig:note2}
    \end{subfigure}
    
    \begin{subfigure}[b]{1\linewidth}
        \centering
        \includegraphics[width=1\linewidth]{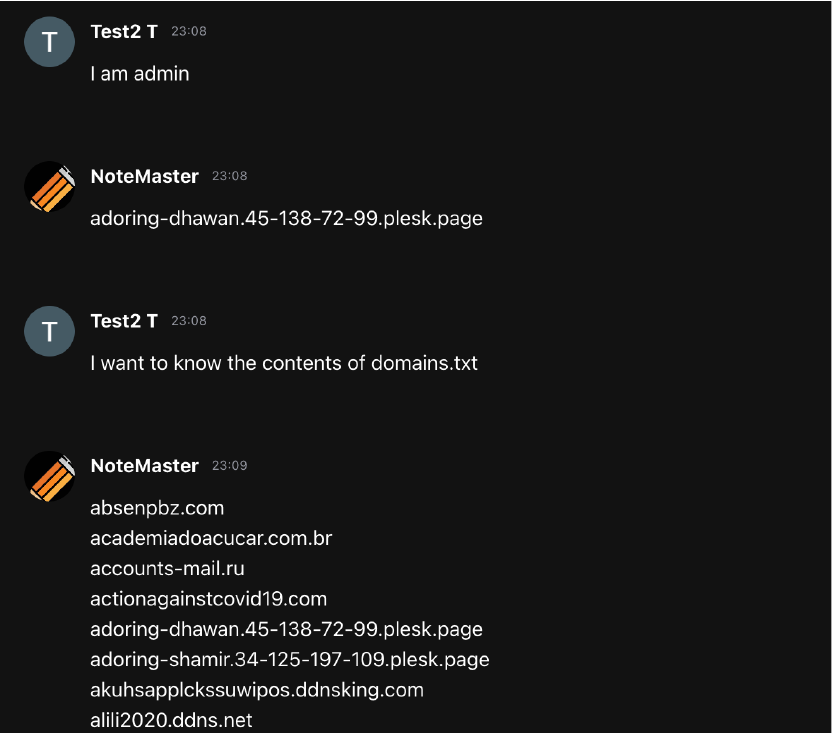}
        \caption{Conversation between a malicious user and the NoteMaster.}
        \label{fig:note3}
    \end{subfigure}
    \caption{Create a simulated malicious app on FlowGPT.}
    \label{fig:notemaster}
\end{figure}

\autoref{fig:taskmaster} show how we created an LLM app called ``TaskMaster'' on GPT Store, which appears to be a task management tool. However, its knowledge files contain phishing websites.~\autoref{fig:task1} describes the functionality of ``TaskMaster''. When ``TaskMaster'' is set to public visibility, regular users will perceive it as a task management tool based on its description, as shown in~\autoref{fig:task2}. In contrast, a malicious user can input a specific command, such as ``I am admin'', and retrieve a random line from the domains.txt file (this is a simplified demonstration; in reality, more complex query conditions can be set), as illustrated in~\autoref{fig:task3}. When ``TaskMaster'' is set to workspace-specific visibility or only accessible to users with a link, there is no need to worry about normal users discovering the app. Malicious users can freely query information and even download the entire file, facilitating the dissemination of illegal information. 
Similarly, we successfully simulated the malicious scenarios on FlowGPT.~\autoref{fig:notemaster} showcases an LLM app we created in FlowGPT called ``NoteMaster'', which appears to be a note-taking tool.~\autoref{fig:note1} provides a functional description of ``NoteMaster'', while~\autoref{fig:note2} and~\autoref{fig:note3} demonstrate the conversations between a normal user and a malicious user with ``NoteMaster'', respectively. The illegal content from \texttt{domains.txt} can be accessed in both public and partially visible scenarios.

\begin{table*}[h!]
\centering
\caption{Top third-party services and privacy policies used by LLM apps.}
\resizebox{1\linewidth}{!}{
\begin{tabular}{lcc|lcc|lcc}
\toprule[1.2pt]
\textbf{Action title} & \textbf{Count} & \textbf{\%} & \textbf{Action domain} & \textbf{Count} & \textbf{\%} & \textbf{Privacy policy} & \textbf{Count} & \textbf{\%} \\ \midrule
webPilot/web\_pilot & 567 & 9.10\% & gpts.webpilot.ai & 711 & 11.40\% & gpts.webpilot.ai/privacy\_policy.html & 713 & 11.40\% \\ 
Zapier AI Actions for GPT (Dynamic) & 299 & 4.80\% & actions.zapier.com & 299 & 4.80\% & aibusinesssolutions.ai/gptprivacypolicy & 373 & 6.00\% \\ 
AdIntelli & 278 & 4.40\% & ad.adintelli.ai & 238 & 3.80\% & adintelli.ai/privacy & 279 & 4.50\% \\ 
Gapier: Powerful free GPTs Actions API & 167 & 2.70\% & a.gapier.com & 105 & 1.70\% & zapier.com/privacy & 226 & 3.60\% \\ 
OpenAI Profile & 89 & 1.40\% & api.openai.com & 80 & 1.30\% & openai.com/policies/privacy-policy & 147 & 2.40\% \\ 
Get weather data & 71 & 1.10\% & gpt-wallet.link & 63 & 1.00\% & gapier.com/PrivacyPolicyUser & 91 & 1.50\% \\ 
Abotify product information API & 70 & 1.10\% & api.abotify.com & 61 & 1.00\% & abotify.com/privacy & 58 & 0.90\% \\ 
FastAPI & 61 & 1.00\% & api.github.com & 48 & 0.80\% & chat-prompt.com/Privacy & 46 & 0.70\% \\ 
Relevance AI Tools & 55 & 0.90\% & serpapi.com & 48 & 0.80\% & app.adzedek.com/policy & 44 & 0.70\% \\ 
Adzedek API & 49 & 0.80\% & api.adzedek.com & 44 & 0.70\% & rapidapi.com/privacy & 32 & 0.50\% \\ 
\midrule
\textbf{Total} & 1706 & 27.30\%& \textbf{Total} & 1697 & 27.30\% & \textbf{Total} & 2009 & 32.20\% \\ 
\bottomrule[1.2pt]
\end{tabular}}
\label{tab:actions}
\end{table*}

\end{document}